\documentclass[]{imag-ms-template}
\usepackage[utf8]{inputenc}
\usepackage[T1]{fontenc}
\usepackage{hyperref}
\usepackage{placeins}
\usepackage[svgnames]{xcolor}
\hypersetup{
  pdfauthor={Pierre-Antoine Comby, Alexandre Vignaud, Philippe Ciuciu},
  pdfkeywords={},
  pdfsubject={},
  pdflang={English},
  colorlinks=true,
  citecolor=blue,
}
\usepackage[american]{babel}
\usepackage{csquotes}
\usepackage{algorithm2e}
\usepackage{xfrac}
\usepackage{bm}

\usepackage[capitalize]{cleveref}
\crefformat{footnote}{#2\footnotemark[#1]#3}

\usepackage{subcaption}
\usepackage{tabularray}
\usepackage[export]{adjustbox}
\UseTblrLibrary{booktabs}

\newcommand{\SNAKE}{\emph{\emph{SNAKE}}}
\newcommand{\TR}{\ensuremath{T\!R}}
\newcommand{\SNRi}{\ensuremath{S\!N\!R_{i}}}
\newcommand{\FA}{\ensuremath{F\!A}}
\newcommand{\TE}{\ensuremath{T\!E}}
\newcommand{\Ts}{\ensuremath{T\sb{2}\sp{*}}}

\usepackage{todonotes}
\setuptodonotes{inline}
\newlength{\todocitewidth}
\newcommand{\todocite}[1]{\settowidth{\todocitewidth}{#1~~~}\todo[color=green!40, inlinewidth=\the\todocitewidth, noinlinepar]{#1}}

\newlength{\logoheight}
\newcommand{\logo}[1]{\raisebox{-0.25\height}{\includegraphics[height=\logoheight]{figs/logos/#1.png}}}

\usepackage[backend=biber]{biblatex}
\author{Pierre-Antoine Comby, Alexandre Vignaud and Philippe Ciuciu}
\date{\today}
\title{SNAKE:\@ A modular realistic fMRI data simulator from the space-time domain to k-space and back}

\begin{document}
\maketitle
\begin{abstract}
  We propose a new, modular, open-source, Python-based 3D+time realistic fMRI data simulation software. \SNAKE or \emph{S}imulator from \emph{N}eurovascular coupling to \emph{A}cquisition of \emph{K}-space data for \emph{E}xploration of fMRI acquisition techniques is the first to simulate the entire chain of fMRI data acquisition, from the spatio-temporal design of evoked brain responses to various 3D sampling strategies of k-space data with multiple coils. We now have the possibility to extend the forward acquisition model to different noise and artifact sources while remaining memory-efficient. Using this in-silico setup, we can provide a realistic and reproducible ground truth for fMRI reconstruction methods in 3D accelerated acquisition settings and explore the influence of critical parameters. This includes the acceleration factor and signal-to-noise ratio~(SNR), on downstream tasks of image reconstruction and statistical analysis of evoked brain activity. 
 In this paper, we present three scenarios of increasing complexity to showcase the flexibility, versatility, and fidelity of \SNAKE: From a temporally fixed full 3D Cartesian to various 3D non-Cartesian sampling patterns, we can compare ---~with reproducibility guarantees~--- how experimental paradigms, acquisition strategies and reconstruction methods contribute and interact together, affecting the downstream statistical analysis.
\newline
\textbf{Keywords:}~\emph{fMRI; Brain Imaging; Accelerated sampling; Compressed Sensing; Simulation; Open Source; Python.}
\end{abstract}

\section{Context and Motivation}

Functional Magnetic Resonance Imaging (fMRI) has emerged as a powerful tool in neuroscience, allowing scientists to investigate human brain function non-invasively.
By measuring $T_2^*$ variations over time due to changes in blood oxygenation and flow, the BOLD effect~\citep{ogawa1990brain}, induced either in response to external stimuli or spontaneously at rest, has provided invaluable insights into cognitive processes, intrinsic functional networks, and brain pathology.
However, conducting task-based fMRI experiments is an expensive and time-consuming endeavor, often requiring access to advanced imaging facilities and substantial expertise in protocol design, data collection, and analysis.

Moreover, the reproducibility and replicability of fMRI experiments and findings has been shown to be a critical issue in cognitive and clinical neuroscience~\citep{liou2006method,griffanti2016challenges,nee2019fmri,bossier2020empirical}.
Reproducibility means that applying the same processing pipeline to the same data should yield the same results regardless of who is running the code, and up to minor changes in parameter tuning.
This challenge has been addressed in the fMRI community by developing and sharing open source software \citep{friston_statistical_2007,smith_advances_2004} and more recently with community initiatives such as fMRIprep~\citep{esteban_fmriprep_2019}, Nipype~\citep{gorgolewski_nipype_2011} and the BIDS format~\citep{gorgolewski_brain_2016}.

In contrast, replicability in fMRI concerns the ability to replicate existing findings on newly collected fMRI data. In that regard, a major roadblock is the sample size, as most of the initial fMRI studies involved small cohorts and are therefore less replicable mainly due to multiple sources of variability between subjects, bias between MR systems, magnetic field strength, etc.~\citep{szucs_empirical_2017,button_power_2013,chen_reproducibility_2018,nakuci_within-subject_2023,marek_reproducible_2022,nee2019fmri,bossier2020empirical}. Moreover, accommodation and habituation effects limit the repeatability of the intra-test \citep{bennett_how_2010}, which severely hinders the comparison of acquisition protocols.

This variability prevents the development of new methodologies to answer the growing needs in the fMRI literature, particularly the quest for higher temporal and spatial resolution to quickly disentangle slow hemodynamic responses~\citep{lewis2016fast,lewis2018stimulus,polimeni2021imaging}, to perform laminar or columnar fMRI imaging~\citep{lawrence2019laminar,yacoub2008high} in an ultra-high magnetic field, or perform an in-depth analysis of a few individuals in multiple experimental paradigms\citep{viessmann_high-resolution_2021,pinho_individual_2020}.

To address these challenges, more complex methods have been developed on the data acquisition and image reconstruction sides.
However, providing a fair comparison between these methods is challenging, due to the lack of existing ground truth data collected on the same individuals but for different imaging setups~(e.g. acquisition parameters, pulse sequences, algorithms for fMRI image reconstruction).

With increasing computing power, the development of (f)MRI simulators has received more attention and interest~\citep{ellis_facilitating_2020,drobnjak_development_2006,erhardt_simtb_2012,welvaert_neurosim_2011}, see also a detailed review available in~\citep{welvaert_review_2014}.
These simulators are designed to provide a ground truth for the fMRI processing pipeline. However, most current simulators are limited to producing (magnitude-only) (f)MRI and do not provide raw k-space data.
Their use can therefore only be seen as a downstream validation of fMRI preprocessing and statistical analysis tools. Furthermore, the critical step of image reconstruction is not considered.

On the other hand, existing k-space data simulators are more dedicated to producing various anatomical MRI scans as a function of multiple input parameters~(pulse sequences parameters, gradients profiles, quantitative maps of tissue properties, etc.)~\citep{stocker_high-performance_2010,jochimsen_efficient_2006,xanthis_mrisimul_2014,kose_blochsolver_2017}, or some approximations~\citep{huang_high-efficient_2023,peretti_generating_2023,petersson_mri_1993} to optimize parameter sequences or readout.
They rely on MR physics simulation models~(i.e.
Bloch equations) and remain computationally expensive.
Their adaptation to artificial fMRI data simulation is beyond their scope as this task would require a prohibitive computing cost for generating the k-space data associated with a single anatomical volume.

In the search for an absolute ground truth, other solutions than simulation have been proposed, such as active phantoms~\citep{cheng_smartphantom_2006,kumar_ground-truth_2021,muresan_automated_2005,brosch_simulation_2002} or fMRI monitoring with other devices, either to reduce the physiological noise and ``clean the BOLD'' effect~\citep{caballero-gaudes_methods_2017}, or to enhance the reconstruction by removing the acquisition artifacts~\citep{amor_impact_2024,duerst_real-time_2015,bollmann_analysis_2017}.
More recently, the development of generative Artificial Intelligence~(AI) has opened new opportunities for the synthesis of realistic MRI data~\citep{pinaya_generative_2023, gopinath_synthetic_2024}.
However, it requires expensive computation that would not scale for generating high-resolution 3D+time fMRI data, and the lack of control and explicability over the yielded sequence of fMRI volumes discards it for validating downstream tasks such as image reconstruction algorithms and statistical analysis.

In general, none of the proposed solutions has succeeded in providing a
\begin{enumerate*}[label=(\roman*)] %
  \item a full control on both the temporally resolved input hemodynamic signal and the output BOLD fMRI time series,
  \item an easy-to-use and reproducible framework through open-source software and
  \item a low computational or operational cost.
\end{enumerate*}

\emph{This paper aims to fill this gap. More precisely, our contribution can be summarized as follows:}

We propose a realistic fMRI simulator, named \SNAKE{} based on an \emph{extended} Fourier model of MRI data acquisition that can create all the required k-space data for evaluating the fMRI processing tool chain: From the definition of an experimental paradigm and the localization of brain activation in a realistic phantom, up to the generation of 3D+time k-space data.
This bottom-up approach gives fMRI scientists a level playing field to explore and compare various properties of acquisition and reconstruction strategies, such as the choice of the sampling pattern (Cartesian vs. non-Cartesian readouts), the acceleration factor, the signal-to-noise ratio~(SNR), etc.
\SNAKE{} could as well be used in the near future to train deep learning models for fMRI image reconstruction, by providing fully parameterized high-quality reference data.

To compare the reconstruction methods in more challenging settings, the ground truth data can be degraded in various ways before and during the acquisition process (motion, static field inhomogeneities, thermal and physiological noise, etc.).

The simulator primarily produces a sequence of k-space volumes sampled with realistic fMRI sequences.
Furthermore, for completeness, we offer the possibility to produce a sequence of fMRI volumes in the image space by plugging in simple but efficient \emph{volume-wise} reconstruction algorithms.
\SNAKE{} has been designed to be as light and as fast as possible to give its users an upper limit to the statistical results that an fMRI experiment can achieve given the acquisition and reconstruction strategies chosen.

In the remainder of this paper, we first propose a review of the available fMRI simulation tools, highlighting their strengths and weaknesses~(\Cref{sec:existing-software}).
Then, we present the underlying model (\Cref{sec:model-equations}) and the design of \SNAKE{} (\Cref{sec:software-desc}).
Finally, we present three typical simulation scenarios (\Cref{sec:scenario-desc}), with basic image reconstruction and statistical analysis (\Cref{sec:scenario-results}), to illustrate the various possibilities of \SNAKE{}, and discuss further the strength and limitations of \Cref{sec:discussion}.

\section{Existing software and their limitations}%
\label{sec:existing-software}

\begin{table*}[h!]
  \caption{Summary of characteristics of different fMRI data simulators.\label{tab:summary-sim}}
  {\footnotesize \begin{center}
  \begin{tblr}{width=0.98\linewidth, colspec={X[0.1, l]X[2.5,l]X[1,l]X[0.5,l,m]X[0.8,l]X[2,l]X[1,l]X[1,l]X[1,l]}, rowsep=2pt, row{Y}={font=\itshape}, row{even}={gray!20},
      cell{3}{1} = {r=6,c=1}{c, white, font=\bfseries},
      cell{9}{1} = {r=5,c=1}{c, white, font=\bfseries} }
  \toprule
 & Simulator Name                                   & {Source}                                                                  & {API}                   & {Sim. Domain}     & Required External Data       & Interface  & Reconstr.                   & Ecosystem               \\
 & {The Virtual Brain\newline\cite{sanz_leon_virtual_2013}}            & \href{https://github.com/the-virtual-brain/tvb-root}{GPL-3.0}             & \logo{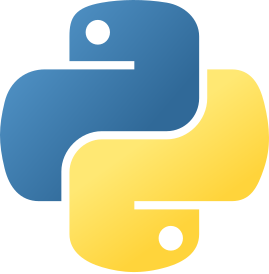}           & Image             &                              & GUI/script & N/A                         &  N/A                    \\
  \midrule \rotatebox{90}{MRI Simulator}
 &{Jemris \newline\cite{stocker_high-performance_2010}}  & \href{https://github.com/JEMRIS/jemris}{GPL-2.0}                          & \logo{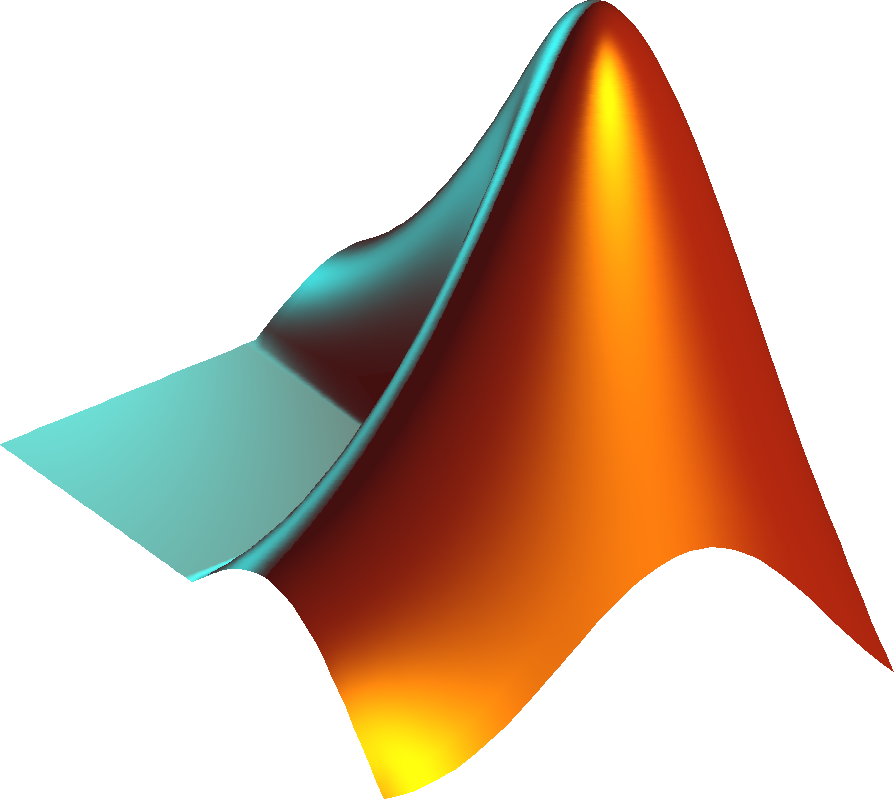}\logo{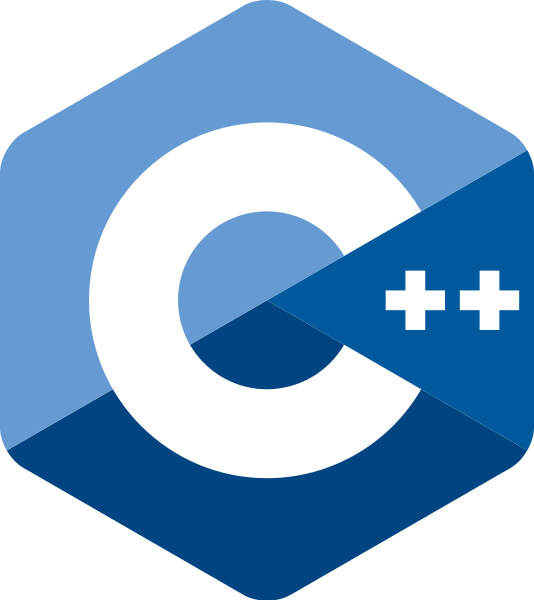} & Bloch             &                              & GUI        & ISMRMD raw data             & N/A                     \\
 &{ODIN \newline\cite{jochimsen_efficient_2006}       }  & \href{https://sourceforge.net/p/od1n/code/HEAD/tree/trunk/odin/}{GPL-2.0} & \logo{cpp}              & Bloch             & Tissue Maps, Sequence        & c++/GUI    & FFT                         & OdinReco                \\
 &{MRILab \newline\cite{liu_fast_2017}                }  & \href{https://github.com/leoliuf/MRiLab}{BSD-2}                           & \logo{matlab}\logo{cpp} & Bloch             & Preset Macros                & GUI        & {FFT\newline Non-Cartesian} & {GUI\newline Gadgetron} \\
 &{Bloch-Solver \newline\cite{kose_blochsolver_2017}  }  & Proprietary                                                               & \logo{python}           & Bloch             & Tissue Maps,                 & script     & FFT                         & N/A                     \\
 &{Fabian \newline\cite{lajous_fetal_2022} }             & \href{https://Medical-Image-Analysis-Laboratory/FaBiAN}{BSD-3}            & \logo{matlab}           & Bloch (EPG)       & Tissue Maps, Sequence        & script     & FFT                         & N/A                     \\
 &{KomaMRI \newline\cite{castillo-passi_komamrijl_2023}} & \href{https://github.com/JuliaHealth/KomaMRI.jl}{MIT}                     & \logo{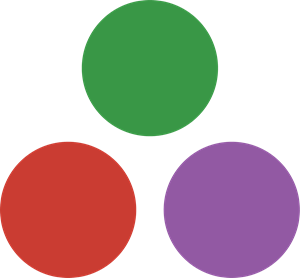}            & Bloch             & Pulseq                       & GUI        & FFT                         &Pulseq                   \\
    \midrule \rotatebox{90}{fMRI Simulator}
 &{POSSUM \newline\cite{drobnjak_development_2006}    }  & \href{https://git.fmrib.ox.ac.uk/fsl/possum}{FSL}                         & \logo{matlab}\logo{cpp} & Bloch             & Tissue Maps Sequence, Events & CLI        & FFT                         & FSL                     \\
 &{Neurolib \newline\cite{cakan_neurolib_2023}        }  & \href{https://github.com/neurolib-dev/neurolib}{MIT}                      & \logo{python}           & Image             & Connectivity Matrices        & script     & N/A                         & Jupyter                 \\
 &{SimTB \newline\cite{erhardt_simtb_2012}            }  & \href{https://github.com/trendscenter/simtb}{Open Source}                 & \logo{matlab}           & Image             & Spatial Maps, Events         & GUI        & N/A                         & MATLAB                  \\
 &{NeuRoSim \newline\cite{welvaert_neurosim_2011}    }   & \href{https://github.com/NeuroStat/neuRosim}{GPL-2.0}                     & \logo{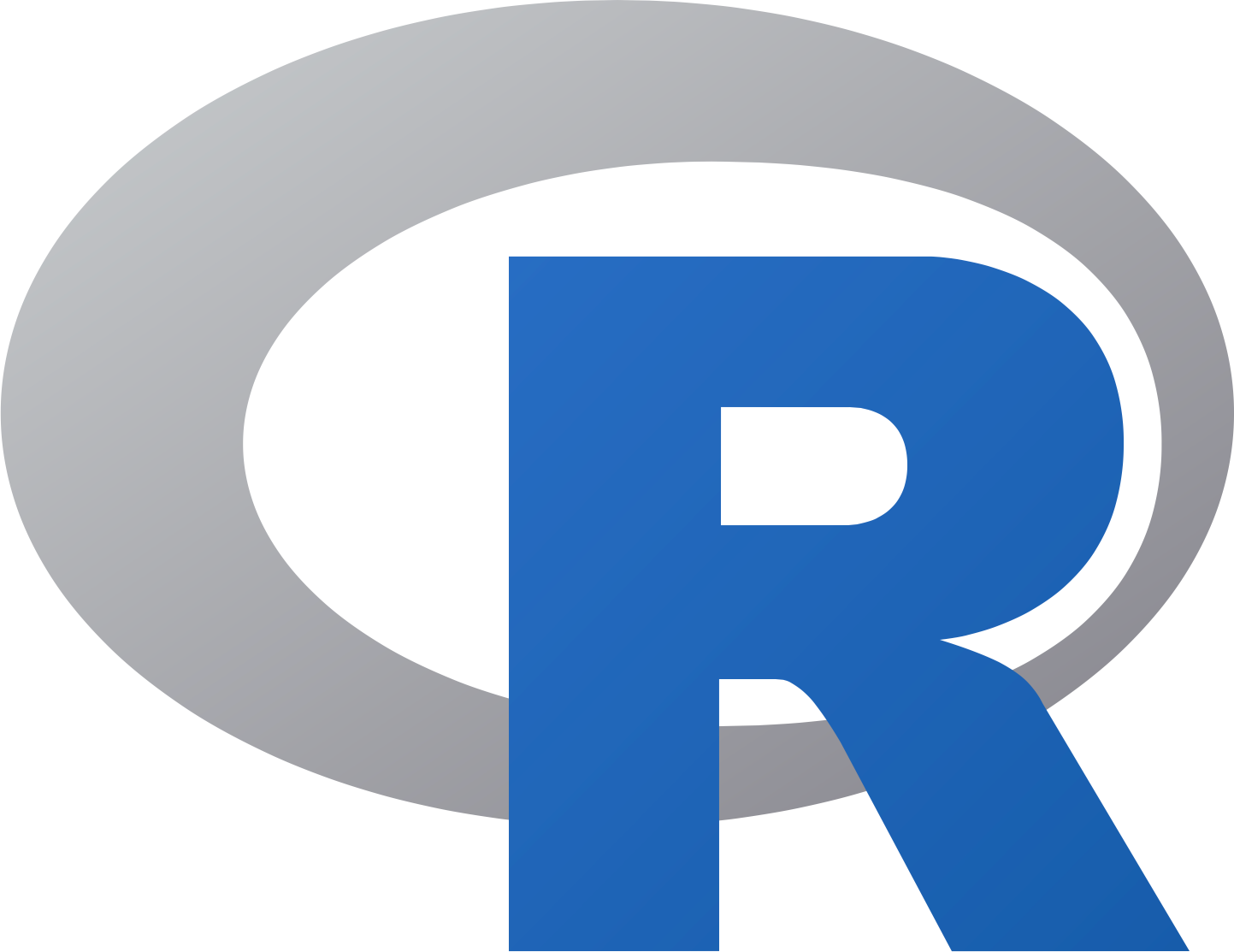}            & Image             &                              & script     & N/A                         & N/A                     \\
 &{fmriSim \newline\cite{ellis_facilitating_2020}     }  & \href{https://github.com/brainiak/brainiak}{Apache-2.0}                   & \logo{python}           & Image             &                              & script     & N/A                         & Brainiak                \\
 &SNAKE-fMRI                                        & \href{https://github.com/paquiteau/simfmri}{MIT}                          & \logo{python}           & Kspace \and Image & Configuration files          & script/CLI & Any (4D methods)            & Any                     \\
    \bottomrule \SetCell[r=1,c=5]{l}
  \end{tblr}\end{center}

API Languages:\hspace{1em} \logo{python} Python \hspace{1em} \logo{matlab} MATLAB\textregistered\hspace{1em}  \logo{cpp} C++ \hspace{1em} \logo{julia} Julia                                                                                                                                 \\
}
\end{table*}

The intricacies of fMRI data create significant obstacles in developing a unified framework for generating comparable datasets.
A detailed examination of the existing literature on fMRI simulation, as highlighted by \cite{welvaert_review_2014}, has shown that the lack of a standardized approach to synthesizing fMRI data severely hinders reproducibility in fMRI research.
This underscores the need for improved transparency in the reporting of experimental design and a more nuanced understanding of the processes involved in the acquisition of fMRI data.
In addition, there exist anatomical MRI simulators based on the Bloch equations that generate contrast-weighted MR images, but they usually remain confined to anatomical imaging with no straightforward extension to fMRI.

In what follows, we provide a broad overview of available MRI and fMRI simulators in \Cref{tab:summary-sim}  and focus hereafter on the main alternative to \SNAKE{}, in increasing order of similarity.

Among existing fMRI data simulation tools, several offer distinct advantages and limitations.
\texttt{The Virtual Brain}~\citep{schirner_brain_2022,sanz_leon_virtual_2013} is an open-source multi-modal brain simulator, focusing on the simulation of brain activity using the structural connectome and interaction between functional networks. It is capable of yielding voxel-wise BOLD fMRI time series. However, it remains complex to master, and is not a dedicated tool for analyzing the acquisition/image reconstruction chain in fMRI.

\texttt{fMRIsim} proposed by \citet{ellis_facilitating_2020}, is a Python package that enables standardized and realistic fMRI time series simulation in the image domain. It was inspired by a parent R package \texttt{neuRosim}~\citep{welvaert_neurosim_2011}.
It allows for the evaluation of complex experimental designs and optimization of the statistical power.
However, it focuses on single-subject simulations and requires manual parameter setting or estimation from real data.
Moreover, it primarily deals with magnitude data and uses additive noise settings, which might restrict its applicability to specific simulations or use cases (e.g. limited resolution).

\texttt{SimTB}, introduced by \citet{erhardt_simtb_2012}, is a MATLAB toolbox specialized in simulating fMRI time series using a separable spatio-temporal model. It offers extensive customization options, including spatial sources, experimental paradigms, tissue-specific properties, noise, and head movement.
\texttt{SimTB} is equipped with both a graphical user interface and scripting capabilities. However, no k-space data is available to assess the performances of various image reconstruction algorithms.

The \texttt{POSSUM} simulator, as outlined by \citet{drobnjak_development_2006}, offers a comprehensive approach to the impact of specific artifacts encountered in the acquisition of fMRI data.
\texttt{POSSUM} accurately simulates these artifacts using Bloch equations and a geometric definition of the brain.
However, its computational cost is high, making simulations of full brains at high resolution prohibitive.
Additionally, it currently only offers the possibility to yield Cartesian data in k-space.
FMRI analysis is deferred to the \texttt{FSL} library, and no comparison with the ground truth is provided as an outcome of the toolbox.

\section{Extended Fourier model for fMRI acquisition}

\label{sec:model-equations}

To bridge the gap between image and k-space simulation, a more realistic forward model is needed. Using the Bloch equation model~\citep{bloch_nuclear_1946} would be prohibitively expensive for whole-brain simulation at high spatial and temporal resolution; instead, we extend the classical model based on the Fourier transform.

\subsection{Single shot acquisition}

Considering the multicoil imaging setup, \(\Ts\) relaxation and off-resonance effects due respectively to signal decay and \(B_0\) inhomogeneities over the field of view (FOV), the signal acquired by the MR system for the \(s\)-th shot in the \(\ell\)-th coil is defined as follows:

\begin{equation}
  \label{eq:fourier_ext}
        y_{\ell,s} (t) = \int_{FOV} m(\bm{r},t)\, \mathcal{S}_\ell(\bm{r})\, e^{-2\imath\pi \Delta f_r(\bm{r})t}\, e^{-2\imath\pi \bm{k}_s(t)\cdot\bm{r}}d\bm{r} %
\end{equation}
where \(\bm{k}_s\) is  the k-space trajectory for the $s$-shot with $k_s(t)$ the k-space location at time \(t\), \(\Delta f_r(\bm{r})\)  the static \(B_0\) inhomogeneity map, \(\mathcal{S}_\ell(\bm{r})\) the sensitivity map for coil \(\ell \in \{1, \dots,L\} \), \(m(\bm{r}, t)\) the base magnetization of the image.
Note that Eq.~\eqref{eq:fourier_ext} holds for both 2D and 3D imaging.
However, the main focus of this paper will be 3D acquisitions, which can proceed to segment the readout either across planes for stacked strategies~(e.g. 3D EPI, stack of spirals)~\citep{poser_three_2010} or across shots for full 3D non-Cartesian strategies~(e.g. SPARKLING or TURBINE, \citet{graedel_ultrahigh_2022}). In the former case, we will use $k_s(t)=[k^x_s(t),k^y_s(t)]$ at a given elevation $z$-th embedded in the definition of $s$, whereas in the latter case we will consider $k_s(t)=[k^x_s(t),k^y_s(t),k^z_s(t)]$.

Assuming steady-state and perfect spoiling regimes, the signal obtains from a Gradient Recall Echo (GRE) sequence parameterized by \((\TR, \TE, \alpha)\), with \(T_1 \gg \TR\), is:
\begin{equation}
  \label{eq:gre_mag}
  m(\bm{r},t) = \rho(\bm{r}) \sin\alpha\, \frac{1-e^{\sfrac{-\TR}{T_1(\bm{r})}}}{1-\cos\alpha\, e^{\sfrac{-\TR}{T_1(\bm{r})}}} \,e^{\sfrac{-t}{\Ts(\bm{r})}}  = \mu(\bm{r}, t_{ref})e^{\sfrac{(t_{ref}-t)}{\Ts(\bm{r})}}
\end{equation}

where \(\rho(\bm{r}), T_{1}(\bm{r}), \Ts(\bm{r})\) are quantitative maps of the proton density, longitudinal~(\(T_1\)) and transverse~(\(\Ts\)) relaxation parameters, and have fixed values during the shot. Different contrasts can be generated, such as the one shown in \Cref{fig:brainweb-phantom}.
\(\mu(\bm{r}, t_{ref})\) describes the signal obtained after the excitation pulse at a point of reference with respect to relaxation \(t_{ref}\), in the case of the in-out sampling pattern, we typically have \(t_{ref}=\TE\), the echo time centered on the $T_{obs}$ long read-out window with \(t \in [-\frac{T_{obs}}{2}, \frac{T_{obs}}{2}]\).

To simplify the computation, the quantitative maps are separated in \(N_{tis}\) tissue types
each with a set of constant MR parameters \(T_1\), \(\Ts\) and \(\rho\):
\begin{equation}
  \label{eq:gre_mag_weights}
  m(\bm{r},t) = \sum_{i=1}^{N_{tis}} w_i(\bm{r})\, \rho_i\, e^{\sfrac{-t}{{T_{2,i}^{*}}}}
\end{equation}

where \(w_i\) is the proportion of tissue \(i\) at voxel location \(\bm{r}\), \(\rho_i\) the proton density and \(T_{2,i}^*\) the \(\Ts\) relaxation parameter of the \(i\)-th tissue type.

Combining \cref{eq:gre_mag,eq:gre_mag_weights,eq:fourier_ext}, we obtain the complete signal equation for a single coil \(\ell\) and a single shot \(s\):

\begin{equation}
    \label{eq:fourier_ext_full}
  \begin{aligned}
    y_{\ell,s} (t) &= \int_{FOV}
                     \sum_{i=1}^{N_{tis}} w_i(\bm{r})\, \rho_i\, e^{\sfrac{-t}{{T_{2,i}^*}}}\,
                     \mathcal{S}_\ell(\bm{r})\, e^{-2\imath\pi \Delta f_r(\bm{r}) t} e^{-2\imath\pi \bm{k}_s(t)\cdot\bm{r}}d\bm{r}\\
                   &= \sum_{i=1}^{N_{tis}} \rho_i\, e^{\sfrac{-t}{{T_{2,i}^*}}} \int_{FOV} w_i(\bm{r})\,
                     \mathcal{S}_\ell(\bm{r})\, e^{-2\imath\pi \Delta f_r(\bm{r}) t} e^{-2\imath\pi \bm{k}_s(t)\cdot\bm{r}}d\bm{r}.\\
  \end{aligned}
\end{equation}

The static off-resonance \(e^{-2\imath\pi \Delta f_r(\bm{r}) t}\) term can be approximated by a sum of separable bilinear terms \(\sum_{p=1}^{P}c_p(t)b_{p}(r) \) \citep{sutton_fast_2003,amor_non-cartesian_2023}.
Furthermore, considering the set of k-space sampling points \(\bm{k}_1, \dots, \bm{k}_n, \dots, \bm{k}_N\) that are acquired possibly off the Cartesian grid and at times \(t_n=(n-1)\Delta t\) (where \(\Delta t\) is typically the dwell time of the scanner, in the order of 10\(\mu\)s), and the spatial locations \((\bm{r}_m)_{m=1}^M \in \mathbb{N}^{3\times M}\) in \(M=N_{x}N_{y}N_{z}\) voxels to cover the FOV, we obtain:

\begin{align} \label{eq:extended_forwardModel}
y_{\ell,s}[t_n] &= \sum_{i=1}^{N_{tis}}\mu_i\, e^{\sfrac{-t_n}{{T_{2,i}^*}}} \sum_{p=1}^P  c_p[t_n]\, \int_{FOV} w_i[\bm{r}]
\mathcal{S}_{\ell}(\bm{r}) b_p(\bm{r})e^{-2\imath\pi \bm{k}_s[t_n]\cdot\bm{r}}d\bm{r} \nonumber \\
y_{\ell,s}[t_n] &= \sum_{i=1}^{N_{tis}}\mu_i\, e^{\sfrac{-t_n}{{T_{2,i}^*}}}  \sum_{p=1}^P c_p[t_n]\, \sum_{m=1}^M w_i[\bm{r}_m]\,
\mathcal{S}_{\ell}[\bm{r}_m]\, b_p[\bm{r}_m] \, e^{-2\imath\pi \bm{k}_s[t_n]\cdot\bm{r}_m}\nonumber \\
  y_{\ell,s}[t_n] &= \sum_{i=1}^{N_{tis}}\mu_i\,e^{\sfrac{-t_n}{{T_{2,i}^*}}}  \sum_{p=1}^P c_p[t_n]\,
                  \mathcal{F}\left\{ b_p\,\mathcal{S}_\ell\,w_i\right\}[\bm{k}_s[t_n]]\, . %
\end{align}

The resulting signal is a doubly weighted sum of Fourier transforms (\(\mathcal{F}\)) first by interpolating coefficients in the k space (\(b_p\)) and second by tissue-specific contrast and \Ts{} decay: \(\mu_i\,e^{\sfrac{-t_n}{{T_{2,i}^*}}}\).
In general, to generate the k space data \(\bm{y}=(y_1, \ldots, y_L)\) in the multicoil array, for a single shot \(s\), the total number of Fourier transform calls in \eqref{eq:extended_forwardModel} is \(N_{tis} P L\).
The computational cost may be reduced by limiting the number of tissues to a few~(e.g. 3 like gray matter, white matter, and cerebrospinal fluid) and / or neglecting \Ts{} relaxation (which is admissible if \(T_{obs}=N\Delta T \ll  \Ts\)) and off-resonance effects. Making all these hypotheses lead to the following \emph{Basic Fourier} model:
\begin{equation}
  \label{eq:fourier-not2}
  y_{\ell,s}[t_n] =  \mathcal{F}\left\{\mathcal{S}_\ell\sum_{i=1}^{N_{tis}}\mu_{i} w_{i}\right\}[\bm{k}_n] =  \mathcal{F}\left\{\mathcal{S}_\ell\,\bm{\mu}\right\}[\bm{k}_s[t_n]]\, . %
\end{equation}
Note that \eqref{eq:fourier-not2} is the model used as the basis for the reconstruction algorithms.
However, neglecting \Ts{} relaxation in fMRI acquisition may yield misleading results, in particular for long-readout trajectories~(such as EVI) or if two neighboring points in the k-space are sampled at the two extremities of the sampling trajectory.

\subsection{BOLD as a TE-sensitive change between shots}

\begin{figure}[htbp]
  \centering
  \includegraphics[width=.7\linewidth]{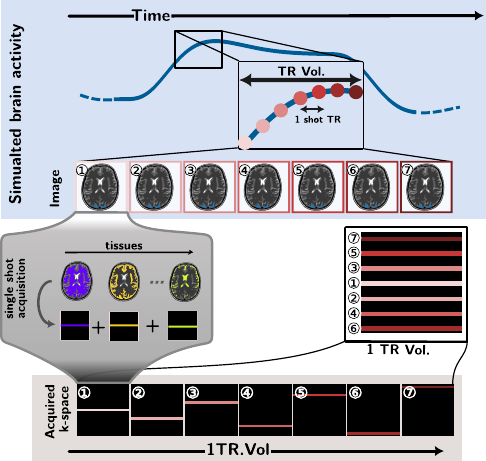}
  \caption{\label{fig:acquisition}%
    Acquisition method implemented in \SNAKE{} -- The case represented is simplified to a 2D Cartesian case (\emph{e.g.} a projected view of a 3D non-accelerated EPI scheme).
    Each shot (i.e. a plane in 3D EPI) of the k-space sampling pattern is acquired separately from an on-the-fly simulated volume in the image domain as shown in the blue frame. The shots are numbered here from  \raisebox{.5pt}{\textcircled{\raisebox{-.9pt} {1}}} to \raisebox{.5pt}{\textcircled{\raisebox{-.9pt} {7}}}.
    The parallel acquisition is performed in parallel for each tissue type to apply the \Ts{} relaxation model \eqref{eq:extended_forwardModel}.}
\end{figure}

Acquisition typically consists of multiple shots, acquired at every \(\TR_{shot}\).
In the context of 3D fMRI, these shots are then grouped together to build a single k-space volume (see \Cref{fig:acquisition}).
Adding functional MRI capability is done by modifying simulation parameters between shots.
In particular, the BOLD effect is modeled as follows:
If we consider that the BOLD effect modifies the baseline gray matter \Ts{}  --- in a given region of interest, and in a simplified manner --- as \(T_{2, BOLD}^* = \frac{1}{R^{*}_{2, GM} + \Delta R_{2}^{*}}\) \citep{jin_source_2006}, we have the following:

\begin{align}
  \label{eq:bold_TE}
  \mu_{BOLD} &= A \cdot \exp({\sfrac{-\TE}{T_{2, BOLD}^*}}) = A \cdot \exp(-\TE \cdot (R_{2, GM}^*+\Delta R_2^*)) \nonumber \\
  \mu_{BOLD} &= \mu_{GM} \exp(-\TE \cdot \Delta R_2^*) \simeq \mu_{GM} (1 - \TE \cdot \Delta R_2^*).
\end{align}
Hence, for every shot \(s\), we can determine the base intensity, in the image domain:
\begin{equation}
  \mu_{BOLD}(t_{s}) = \left(1 - \TE\cdot\Delta R_2^*  \tilde{h}(t_{s})\right) \cdot \mu_{GM}(t_{s})\,, %
\end{equation}

where \(\tilde{h}(t)\) is the normalized hemodynamic response such that \(\max \{\tilde{h}(t)\} = 1\).
In the context of task-based fMRI, it is modeled by convolving a sequence of events~(event-related paradigms) or blocks~(block paradigms) with a reference hemodynamic response function HRF \citep{glover_deconvolution_1999,ciuciu2003unsupervised}.
In our simulation, we used \(\Delta R_{2}^{*} = -1\)Hz (following the value used by \citet{jin_source_2006}), which generates a 2. 5\% increase in the BOLD contrast at \(\TE = 25\)ms.

\subsection{Noise and SNR calibration}
\label{sec:noise-snr-calibration}

The (f)MRI signal in the k-space is also corrupted by thermal noise sources that arise from the acquisition process in two forms: Brownian motion of spins and random fluctuation in the RF receiver processing chain.

To model those effects, we add a complex multivariate Gaussian noise over the coil component $\ell \in (1,\dots,L)$ for every shot $s$.
An existing coil covariance matrix $\bm{\Sigma} \in \mathbb{C}^{L\times L}$ can be supplied to match an existing hardware set-up.
Moreover, the global noise variance can be tuned manually, to set the \emph{input SNR} in k-space.
Concretely, the noise for each time point in the shot is sampled from $(n_{\ell,s}[t_n])_{1\le\ell\le L} \sim \mathcal{N}(0, \frac{E(\hat{x})}{\SNRi}\bm{\Sigma})$
where \(\bm{\Sigma}\) is the coil covariance matrix.  \(E(\hat{x})\) is the energy of the \emph{ideal} phantom, acquired at \TE{}, and \SNRi{} is the input SNR defined by the user in k-space. To add noise to a full shot, we draw $N$ realizations of the $L$-dimensional noise vector \((n_{\ell,s}[t_n])_{\begin{subarray}{l} 1\le\ell\le L \\ 1\le n \le N\end{subarray}}\).
Then noisy data is formed as follows:
\begin{equation}
  \label{eq:snr}
\forall t_n \in 1,\ldots, N, \quad \tilde{y}_{\ell,s}[t_n] = y_{\ell,s}[t_n] + n_{\ell,s}[t_n] \, .
\end{equation}

It is possible to calibrate the value of \SNRi{} using experimental data, by computing the energy ratio of k-space shots collected with and without RF excitation.

Another solution is to view the joint system of acquisition and reconstruction together, and consider a known case of image quality output. In this case, the value of \SNRi{} is tuned to obtain the desired image quality, gathering all missing modeling aspects in this Gaussian noise.

More structured noise sources~(e.g.
physiological noise such as heartbeat, breathing rate, motion) could be superimposed on the simulated signal.
However, the wide variety of options in this field for modeling such noise components is beyond the scope of this article.
However, \SNAKE{} already offers a wide flexibility through the handler mechanism (see \cref{sec:modularity}).
Future work (open to contributions) will go into more detail on the implementation of these noise sources.

\subsection{Summary of general hypotheses on the acquisition model}
\label{sec:summary-hypotheses}

In general, the current capabilities of SNAKE are currently restricted by the following hypotheses.
\begin{enumerate}[label=(\roman*), noitemsep]
  \item\label{hyp:tissue} Tissue parameters and physiology are frozen during the acquisition of a single shot: \(T_{1}\), \Ts, \(\rho\), \(\chi\) are constant when the signal \eqref{eq:extended_forwardModel} is computed.
  \item\label{hyp:BOLD1}The BOLD effect is linearly sensitive to \TE.
  \item\label{hyp:BOLD2} The BOLD effect does not modify the phase of the signal.
  \item\label{hyp:noise} Complex valued Gaussian noise is added in the k-space across coils (modeling the thermal noise).
The noise level is calculated from a user-prescribed SNR.
  \item The acquisition of shots follows \Cref{eq:extended_forwardModel}. As \Cref{sec:scenario-results} will demonstrate, this model can be simplified to \eqref{eq:fourier-not2} if the trajectories do not have long readouts and are temporally smooth.
\end{enumerate}

These are the minimal hypotheses on which the implementation of \SNAKE{} is based. However, it also provides a flexible framework for adding more complex models (e.g. motion or other physiological perturbations) through the handler mechanism as described in \cref{sec:modularity}.
To make \SNAKE{} user-friendly and reduce its parameterization, additional hypotheses can be formulated.
However, the potential bias introduced by these simplifications should be carefully considered. In \Cref{sec:work-hypotheses} we describe further restricting hypotheses for our study case.

It should be outlined that \SNAKE{} does not aim to reproduce the full complexity of MR physics, but instead to provide a realistic and reproducible framework that allows us to simulate various acquisition scenarios in fMRI with full control over the ground truth and its parameterization.
As such, \SNAKE{} provides an upper limit on the statistical sensitivity/specificity compromise given the definition of an acquisition and image reconstruction set-up.
Hence, \SNAKE{} is defined as an instrumental framework for benchmarking fMRI reconstruction methods that comes with reconstruction methods and statistical analysis tools to perform end-to-end validation of fMRI acquisition and reconstruction methods, as depicted in \Cref{fig:block-chain}.

\section{Main characteristics of \SNAKE{} implementation}
\label{sec:software-desc}

\SNAKE{} has been designed as a fully reproducible modular fMRI simulator capable of generating k-space data efficiently.
In what follows, we give a broad overview of the main features of \SNAKE{}.
Then they will be illustrated in \Cref{sec:scenario-desc} and \Cref{sec:scenario-results}.

\subsection{Modular Approach}
\label{sec:modularity}
\SNAKE{} adopts a modular approach to simulate 3D + time fMRI data.
\begin{figure*}
  \centering
  \includegraphics[width=.9\linewidth]{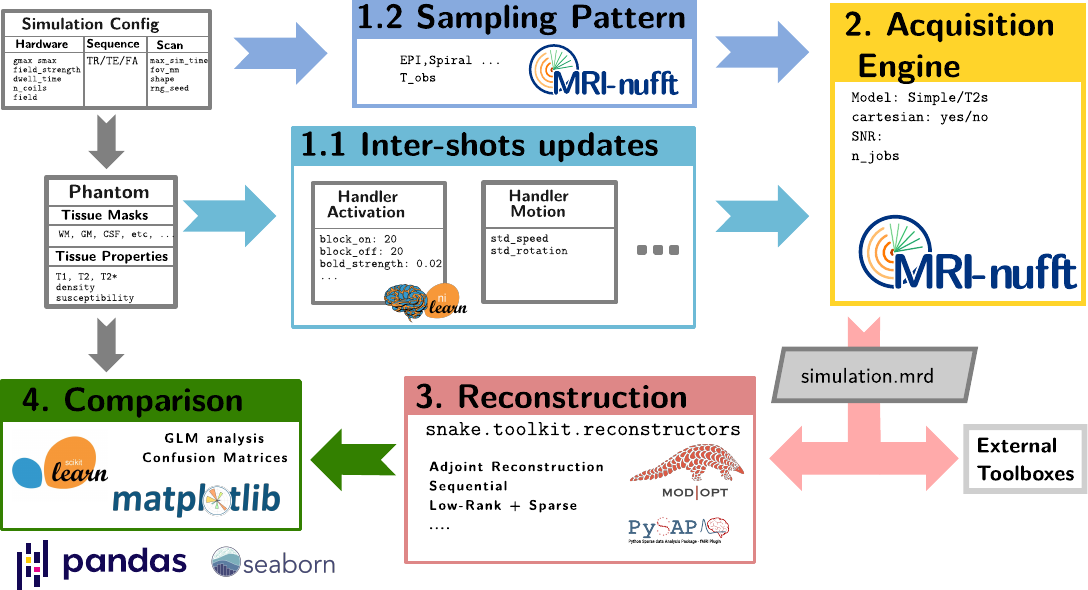}
  \caption[handlers-chaining]{\label{fig:block-chain}Modular design of the \SNAKE{} simulator, which embeds an ecosystem of Python packages for its different building blocks.
    \begin{enumerate*}
      \item The dynamics of the acquisition are simulated, along with the sampling pattern for each shots.
      \item The acquisition engine gathers the multiple shots involved in the acquisition as well as the current state of the 3D anatomical phantom. It relies on the \texttt{MRI-nufft} package.
      \item Then the output k-space data is reconstructed into the image domain using some specific method implemented in \texttt{pysap-fmri}/\texttt{ModOpt}\citep{farrens_pysap_2020}, thereby producing an estimation of the fMRI time-series.
      \item Finally, the sequence of fMRI volumes is analyzed through a general linear model~(GLM) defined from the experimental paradigm and implemented in \texttt{nilearn}\citep{abraham_machine_2014}. Statistical maps are produced and compared to the ground truth used at the simulation stage using confusion matrices and statistical metrics~(accuracy or ACC, balanced accuracy or BACC).
    \end{enumerate*}
  }
\end{figure*}

Typically, the simulation begins with the definition of an anatomical phantom of the brain in the image domain~(see \Cref{fig:brainweb-phantom} in \Cref{sec:brainweb-phantom}), and proceeds to add BOLD contrast and various noise sources through object called ``handlers'' before generating the k-space data from a user-defined sampling pattern.
As shown in \Cref{fig:block-chain} those handlers can be chained to produce complex behaviors from simple operations. Moreover, numerous sampling patterns (especially non-Cartesian ones) can be generated through the use of the \texttt{MRI-NUFFT} library. The acquisition process is depicted in \Cref{fig:acquisition}. \SNAKE{} supports both \eqref{eq:extended_forwardModel} and \eqref{eq:fourier-not2} through dedicated ``engines'' that are optimized for parallel computing of shots on GPU.

This option allows users to assess the need for a more complex and possibly more computationally demanding model in the context of the chosen scenario.

Moreover, \SNAKE{} directly provides access to variational reconstruction methods using \texttt{PySAP-fMRI}\footnote{\url{https://github.com/paquiteau/pysap-fmri}} and \texttt{ModOpt}~\citep{farrens_pysap_2020}.

At the end of the processing chain, it is possible to compare the reconstructed fMRI images and time series data with ground-truth simulation, and evaluate the effect of acquisition parameters (SNR, acceleration factor, etc.) and reconstruction strategies~(density compensated adjoint Fourier, compressed sensing reconstruction, etc.) on image quality, as well as on statistical sensitivity/specificity compromise. Statistical analysis is performed using the \texttt{nilearn} package~\citep{abraham_machine_2014}.

\subsection{Performance,  reproducibility and scalability made easier for neuroscientists}

\subsubsection{Performance}
\label{sec:performance}
Storing in RAM the full high spatiotemporal simulation is challenging at the high spatial and temporal resolution (340 Gigabytes are required for a 1mm-iso volume, with a unitary TR of 50ms shot-wise for 5 minutes of a typical fMRI run) and each substantial change, like adding noise would create a new copy of this data.
Instead, we propose to yield the data to be acquired shot-wise on the fly, as each time point in the time series can be computed from a sequence of transformations applied to a single anatomical volume (for instance, adding the BOLD contrast, noise, or using motion parameters).
Moreover, whenever possible, the computations are performed on GPU and shot-wise acquisitions are performed in parallel, the data is also eagerly off-loaded to hard disk when it is no longer required for computations. The k-space data generated by SNAKE is exported in the standardized ISMRMRD format (\texttt{.mrd}, \citet{inati_ismrm_2016}).

\subsubsection{Reproducibility}

Enabling reproducibility in the study of fMRI processing methods and their benchmarking is at the heart of the development of \SNAKE{}.
This simulator can be installed directly from the Python package archive~(\url{https://pypi.org/project/snake-fmri/}) and its core only depends on standard and well-tested Python packages.
Simulation setup can be shared through \texttt{.yaml} files describing the recipe for building a simulation\footnote{\label{fn:config-repo}The detail configuration and run scripts for scenarios of \Cref{sec:scenario-desc} are available at \url{https://github.com/paquiteau/snake-fmri}}.

\subsubsection{Scalability and interoperatibility}

Using the \texttt{.mrd} file as output, we open the door for interoperability of \SNAKE{} with other toolboxes for image reconstruction, such as SigPy~\citep{ong_frank_sigpy_2019} or BART~\citep{Uecker_Ong_Tamir_2015}.
We also provide optimized data-loader for the \texttt{.mrd} files generated by \SNAKE{}.

Furthermore, based on the \texttt{hydra} framework~\citep{yadan_hydra_2019} we can run multiple simulations with different parameters or handlers and to perform image reconstruction of fMRI volumes and simple  statistical data analysis to compare competing approaches. It also allows us to scale up simulations from a single laptop to high-performance computing clusters.

\section{Numerical experiments}
\label{sec:scenario-desc}

In this section, taking advantage of the modularity and scalability of \SNAKE{}, we demonstrate the use of a controlled simulation framework to explore the potential benefits and challenges of moving to higher resolution in space and time for fMRI experiments.
Results and analysis are presented in the next section.

\subsection{Acquisition scenarios}

First, we detail three simulated scenarios of increasing complexity (from 3D Cartesian low spatial resolution to 3D non-Cartesian high temporal or spatial resolution) with basic reconstruction and statistical analysis pipeline.
All scenarios simulate a five-minute-long run with full-brain coverage during a simple visual stimulation using a standard block-design paradigm at 7T, which alternates 20s-on and 20s-off periods. In response to visual stimuli, we induce a 2. 5\% change in BOLD contrast (following~\Cref{eq:bold_TE}) in a region of interest (ROI) in the occipital cortex. This ROI is defined from a fuzzy segmentation of gray matter that intersects an ellipse located in the occipital cortex.

Furthermore, we ran both the \Ts{} relaxation model~\eqref{eq:extended_forwardModel} and the simplified Fourier model~\eqref{eq:fourier-not2} for the acquisition, to determine in which settings a more complicated model is required.

\subsubsection{Working hypotheses to simplify acquisition}
\label{sec:work-hypotheses}

In all scenarios, the acquisition model considers only the intrinsic phenomenon of MR physics and lets aside all other modeling steps~(motion, physiological noise, off-resonance effects) that could be compensated using a more complicated reconstruction setup. In particular, in addition to the fundamental hypothesis in \Cref{sec:summary-hypotheses}, we assume that:

\begin{itemize}[noitemsep]
  \item Following Hypothesis~\ref{hyp:noise} in \Cref{sec:summary-hypotheses}, the coil covariance matrix \(\bm{\Sigma}\) is set to identity, and we use a user-defined SNR level (\(\SNRi=1000\), determined using the calibration step described in \Cref{sec:noise-snr-calibration}).

  \item \emph{There is neither motion nor physiological perturbation (aside from the BOLD signal)}: in the context of acquisition and reconstruction benchmarking, we assume that the motion could be compensated, even though this task may require in practice navigator echoes or dedicated hardware.
  
  \item \emph{Off-resonance effects are not considered}:
  From this point of view, we put ourselves in an ideal situation with minimal static and dynamic \(\Delta B_{0}\) inhomogeneities. Furthermore, we recently showed that it can be compensated using field cameras, in addition to a static map \(\Delta B_{0}\) in the forward model at reconstruction time.~(cf.\citep{amor_impact_2024,amor_non-cartesian_2023}) \eqref{eq:extended_forwardModel}.

  \item \emph{Reducing the number of tissues}: Limiting ourselves to the three cortical tissues (white matter (WM), gray matter (GM), cerebrospinal fluid (CSF)) does not hinder the comparison between ground-truth and reconstructed images or statistical analysis.
\end{itemize}

Such assumptions allow us to focus on the core components of \SNAKE{} (See \Cref{sec:summary-hypotheses}), reduce the computational load,  but nonetheless provide an upper-bound on the performances of the tested scenarios. Possible extensions can be added by implementing the appropriate handlers.

The three scenarios and their computational cost are summarized in \Cref{tab:scenarios-desc}. Details of all parameters used are available in the \texttt{.yaml} configuration files in the \SNAKE{} source repository\cref{fn:config-repo}.

As the unitary $\TR_{shot}$ and the simulation time (5 minutes) are the same for all scenarios, each simulation has a budget of 6000 shots to acquire. Depending on the number of shots allocated per frame (for a specific acceleration factor AF), we end up with a variable number of k-space volumes across the three scenarios.

The simulation times reported in \Cref{tab:summary-sim} underline the additional overhead of using \eqref{eq:extended_forwardModel} instead of \eqref{eq:fourier-not2} for the simulation. In Scenario 1 the simulation time is limited by I/O communications, whereas Scenario 2 and 3 are computationally bounded.

\begin{table}[!htb]
  \SetTblrStyle{caption}{hang=0pt,halign=c}
  \centering
  \begin{talltblr}[
    caption={\label{tab:scenarios-desc}Overview of simulated scenarios and their computation requirements -- \(L, N_{s}, n_{jobs}\)  specify the number of simulated coils, the number of shots acquired to create a k-space volume and the number of concurrently simulated shots, respectively.},
    remark{Hardware}={CPU:~Intel i9-13900H, RAM: 32Go, GPU:~NVIDIA RTX 2000 Ada.},
    remark{Common sequences parameters}={\(\TR_{shot}=50\)ms, \(\TE=25\)ms, \(\FA=12^{\circ}\).},
    ]{colspec={|l|llrrrrrr|rr|}, colsep=3pt,
       cell{1}{1-X}={r=2,c=1}{c},
       cell{1}{Y}={r=1,c=2}{c},
     }
     \toprule
     Setup                                     & Res. & Readout       &  \(SNR_{i}\) &  \(L\) &\(N_{s}\) & \(T_{obs}\) & \(\TR_{vol}\) & \(n_{jobs}\)  & Time          &          \\
                                               &      & Trajectory     &              &            &            &                &               &               & Fourier \eqref{eq:fourier-not2} & With \Ts{} \eqref{eq:extended_forwardModel} \\
     \midrule
     \hyperref[sec:scenario1-desc]{Scenario 1} & 3mm  &EPI             &  1000        &  1         &  44        &   25ms         &2.2s           &             6 & 3min54sec        & 3min55sec   \\
     \hyperref[sec:scenario2-desc]{Scenario 2} & 3mm  &SoS &  1000        &  8         &  14        &   30ms         &0.7s           &             6 & 1min31sec        & 5min10sec   \\
     \hyperref[sec:scenario3-desc]{Scenario 3} & 1mm  &SPARKLING       &    30        &  32        &  48        &   25ms         &2.4s           &             3 & 1h34min20sec     & 3h58min40sec  \\
     \bottomrule
   \end{talltblr}
\end{table}

\subsubsection{Scenario S1: 3D fully sampled Cartesian readout}
\label{sec:scenario1-desc}

As a first validation example, we simulated the acquisition of 3D Echo Planar Imaging~(EPI) data, as an implementation of a 3D Cartesian readout.
The 3D EPI is segmented plane by plane in the k space (as in \citet{poser_three_2010}).
Each slice was fully sampled at a 3mm isotropic resolution (matrix size: \(60 \times 71 \times 60\)), with a \(\TR_{shot}=50\)ms.
The volume-wise temporal resolution is \(\TR_{vol}=2.2\)s at Ernst flip angle~($12^\circ$), leading to a \SNRi{} of 1000 calibrated on real acquisitions (cf. \Cref{sec:noise-snr-calibration}).
Since the data were collected at the Nyquist rate, we restricted ourselves to a single-coil acquisition.
This simple configuration can be simulated on a standard laptop in a few minutes, a similar configuration acquired with POSSUM would have taken several hours, for fewer slices (see~\citet{drobnjak_development_2006}, Table 2), due to its more mathematically involved model related to MR physics.

\subsubsection{Scenario S2: 3D under-sampled stack of spirals~(SoS) readout with VDS acceleration along the stacking axis}%
\label{sec:scenario2-desc} %

\begin{figure}[!hbt] \centering
  \begin{subfigure}[t]{0.6\linewidth}
        \captionsetup{width=.8\linewidth}
    \includegraphics[width=\textwidth]{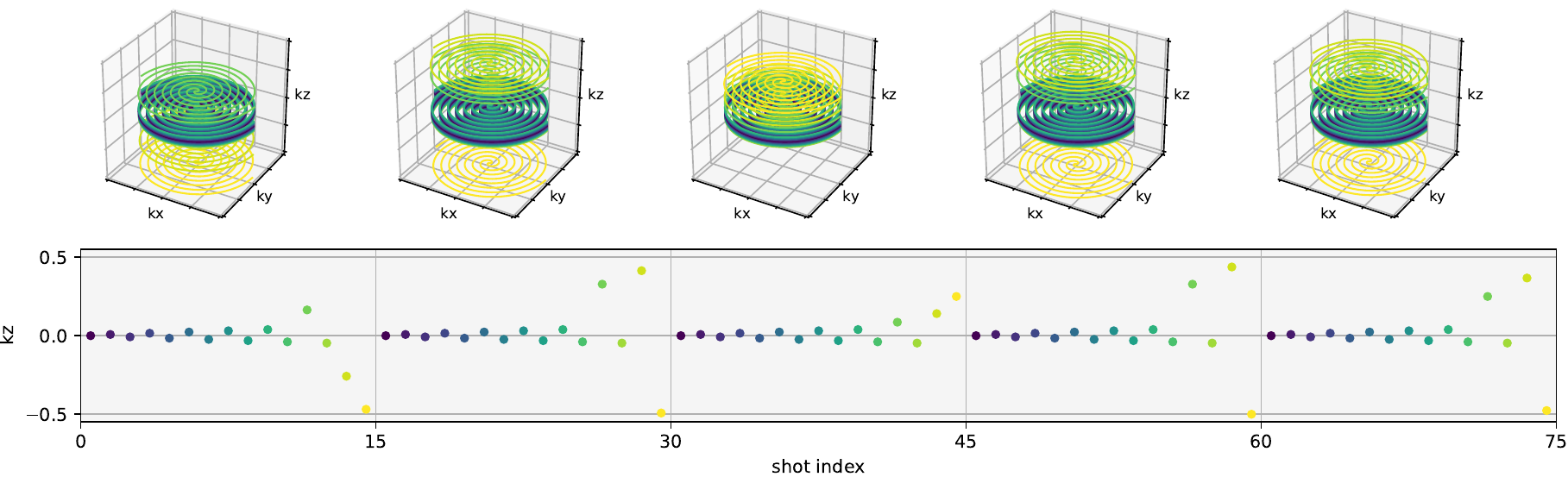}
    \caption{\label{fig:scenario2-traj} Dynamic stack of spiral trajectory used in Scenario 2, with a center-out order for acquiring shots in the stacking.}
  \end{subfigure}\hfill%
  \begin{subfigure}[t]{0.4\linewidth}
    \captionsetup{width=.8\linewidth}
    \centering
    \includegraphics[width=0.6\textwidth]{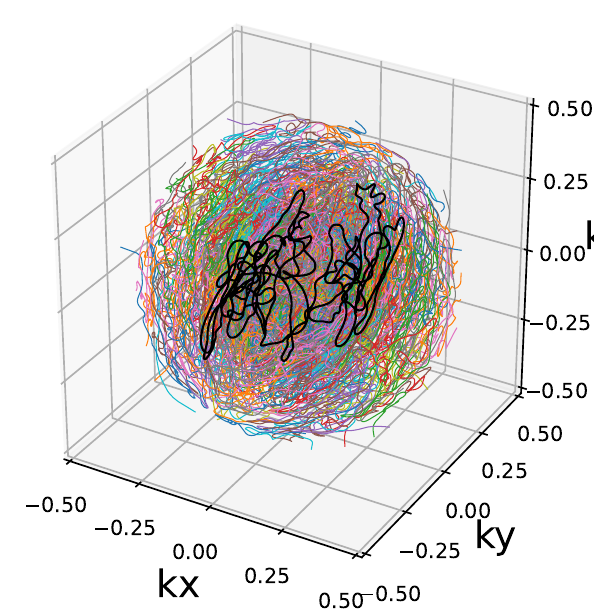}
    \caption{3D SPARKLING trajectory used in Scenario 3: 48 shots, with \(TR_{shot}\)=50ms.}
  \end{subfigure}
  \caption{Non-Cartesian k-space sampling trajectories used in Scenarios 2 and 3.}
\end{figure}

The second scenario explored the possibility of \SNAKE{} for accelerated imaging based on Compressed Sensing~(CS) for data acquisition and image reconstruction.
The resolution and FOV remained the same as for Scenario S1, but the goal was to increase the temporal resolution.
To do so, we performed an acceleration on the 2D plane (\(k_{x},k_{y}\)) using an in-out spiral acquisition and second we implemented a spiral stack using a variable density sampling along \(k_{z}\), i.e. the stacking dimension~(see parameters in \Cref{tab:scenarios-desc}). This sampling evolved across scans: Around 10\% of the center planes were constantly acquired, while we used an acceleration factor AF=4 on the outer planes, as shown in \Cref{fig:scenario2-traj}.
We eventually collected 14 spiral shots per volume, reaching in general \(\TR_{vol}=0.7s\) as volume temporal resolution ($\TR_{vol}$).
To compensate for aliasing artifacts, we simulated a multicoil acquisition with $L=8$ receiver coils.
Using GPU-accelerated NUFFT, the simulation time of the acquisition at 3mm isotropic resolution took around 1.5 minutes of computation (\Cref{tab:scenarios-desc}) in the
CPU-based NUFFT implementation, which is also available, but remains slower.

This approach has already been studied numerically and experimentally in \citep{petrov_improving_2017} with image reconstruction strategies that take advantage of this acceleration mechanism, such as low rank + sparse regularization~\citep{petrov_improving_2017, lin_efficient_2019, otazo_low-rank_2015, chiew_recovering_2018}. Here, we limit image reconstruction strategies to frame-wise approaches, described in \Cref{sec:scenario-reconstruction}.

\subsubsection{Scenario S3: Fully 3D SPARKLING under-sampled readout}
\label{sec:scenario3-desc}

We finally simulated a third scenario using a fully 3D non-Cartesian sampling pattern, based on 3D SPARKLING acquisition~\citep{chaithya_optimizing_2022,amor2024non} over $L=32$ coils at an isotropic resolution of 1mm and $TR_{vol} = 2.4$ s. 3D SPARKLING implements a variable density sampling according to a prescribed sampling density in k-space, while complying with the hardware constraints on gradient magnitude $G_{\rm max}$ and slew rate $S_{\rm max}$. These values are user-defined and were set to $G_{\rm max}=40$mT/m and $S_{\rm max}=180$T/m/s. The target density was a radially decaying distribution parameterized by a cut-off and decay parameters set to $(C,D)=(0.25,2)$, see~\citep[Eq. (3)]{amor2024non} for details.
Fully 3D SPARKLING allows us to further accelerate the acquisition process compared to Scenario~2 to reach higher spatial resolution.
As we move to higher spatial resolution compared to scenarios S1 and S2 (voxels are 27\(\times\) smaller), we multiplied the \SNRi{} level by a 30-fold factor.

The parameters for this scenario are based on the experimental setup of~\citep{amor2024non}, which is described in \Cref{sec:scenario-reconstruction}, and in \Cref{fig:scenario2-recos}, the upper row.
In contrast to Scenario~2, here we adopted the ``scan and repeat mode'' that consists of sampling the same k-space locations across consecutive frames. Time-varying 4D SPARKLING acquisition for fMRI is left for future work and currently beyond the scope of \SNAKE{}.

\subsection{Reconstruction Strategies for Scenario 2 \& 3}
\label{sec:scenario-reconstruction}

For simplicity, as the primary focus here was on the simulation of k-space data, we restrict ourselves to a classical, frame-wise CS based image reconstruction with a standard sparsity-enforcing regularization term in the wavelet domain:

\begin{equation}
  \label{eq:sequential}
  \widehat{\bm{x}}_{t} =  \arg\min_{\bm{x}\in \mathbb{C}^{M}} \frac{1}{2} \sum_{\ell=1}^{L}\left\|\bm{\mathcal{F}}_{t} \mathcal{S}_{\ell}\bm{x} - \bm{y}_{t, \ell} \right\|_{2}^{2} + \mu_t g(\bm{\Psi}\bm{x}) \, .
\end{equation}
This means that each frame \(t\) is reconstructed independently of the others, as we use the following notation:
\begin{itemize}[noitemsep]
  \item \(\mathcal{F}_{t}\) is the Fourier transform operator for the trajectory at time \(t\).
  \item \(\bm{\Psi}\) is a orthogonal wavelet transform such as \texttt{sym-8}.
  \item $\mu_t>0$ is the regularization parameter for frame \(t\).
  \item \( g(\cdot)\) is the regularization function, here a standard \(\ell_{1}\)-norm is used.
\end{itemize}
\noindent The cost function to be minimized is convex but non-smooth. Its global minimizer can be found iteratively using a wide range of proximal gradient methods with possible acceleration schemes such as FISTA, POGM~\citep{beck_fast_2009,kim_adaptive_2018}. As illustrated in \Cref{fig:block-chain}, CS-based image reconstruction was performed using the PySAP library~\citep{farrens_pysap_2020}, and the implementation of the POGM algorithm~\citep{kim_adaptive_2018}. Specifically for fMRI we implemented a dedicated plugin called \texttt{pysap-fmri}.
To reduce the number of free parameters, the regularization parameter for each frame \(\mu_t\) was estimated using Stein's unbiased risk estimate~(SURE) principle~\citep{donoho_threshold_1994}, as detailed in \Cref{sec:app_SURE}. %

\begin{figure}[hbtp]
    \centering
    \includegraphics[width=0.5\textwidth]{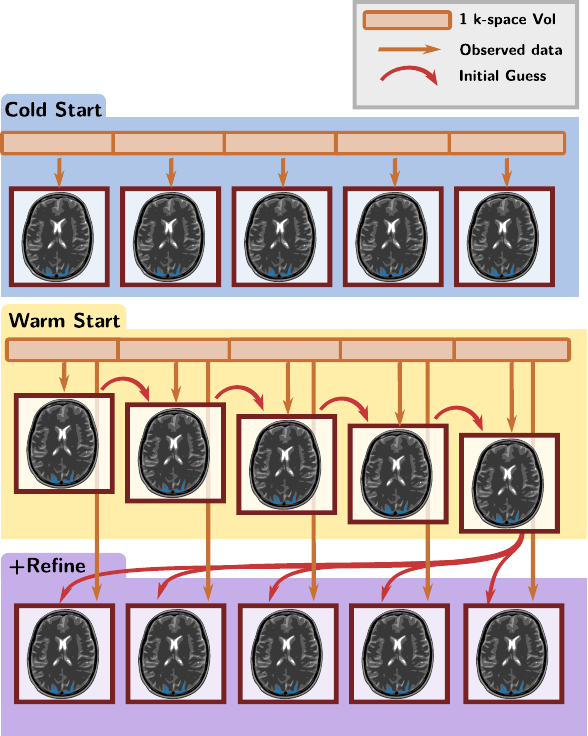}
    \caption{\label{fig:scenario2-recos} Different methodologies for sequential reconstruction used in Scenario S2. Top: \emph{cold start} reconstruction, each frame is reconstructed independently. Center: \emph{warm start} reconstruction, each frame is reconstructed using the previous frame as initialization. Bottom: \emph{refined} reconstruction, after a warm start reconstruction, the last frame is used as initialization for all other frames.}
  \end{figure}

The versatility of \SNAKE{} allowed us to investigate two different acquisition strategies in the space k within scenario S2, the first based on \emph{static} SoS and a second one associated with a \emph{dynamic} SoS. In the static regime, a constant spiral stack was used for all frames, whereas in the dynamic regime, a spiral stack that varied over time was designed by picking up $k_z$ plans randomly for each and every frame, as shown in \Cref{fig:scenario2-traj}.
Then CS reconstruction of each frame, as described in \Cref{fig:scenario2-recos}, was carried out according to three different mechanisms:

\begin{enumerate}[label=(\roman*),noitemsep]
  \item in a \emph{cold-start} manner, where each volume in frame $t$ is reconstructed by solving \eqref{eq:sequential}
  independently of previous frames $\{1, \ldots, t-1\}$, as illustrated in \Cref{fig:scenario2-recos}[top].

  \item in a \emph{warm-start} manner, where the volume reconstructed in frame \(t\) by solving \eqref{eq:sequential} is used as initialization to reconstruct the following volume frame \(t+1\), as shown in \Cref{fig:scenario2-recos}[center].

  \item in a \emph{smart} manner, using an additional \emph{refined} initialization, where the last volume reconstructed in frame $T$ using the previous warm-start strategy was eventually injected as a new set-up for all previous frames, as explained in \Cref{fig:scenario2-recos}[bottom]. Coupled with dynamic SoS at acquisition, this approach is instrumental in visiting all k-space measurements across the consecutive frames.
\end{enumerate}

All these variations are based on a frame-wise 3D image reconstruction strategy and, as such, provide a memory-efficient implementation. However, they do not leverage all 4D fMRI k-space data at once, in contrast to low-rank+sparse methods~\citep{petrov_improving_2017,otazo_low-rank_2015}.

Estimation of \(\mu_t\) in each frame (detailed in \Cref{sec:app_SURE}) plays a critical role in the performance of the \emph{warm} and \emph{refined} strategies. The first reconstructed frames are highly regularized, but as we progress towards the end of the run, the estimates of \(\mu_t\) using the SURE-based methods get smaller, as we progressively embed more information to reconstruct volume \(x_t\).

\subsection{Evaluation methods}

To evaluate the performance of each scenario and the impact of data acquisition and image reconstruction strategies, each combination was submitted to a standard fMRI statistical analysis pipeline, which consists of applying a general linear model (GLM) and testing the positivity of the single modeled experimental condition~(visual) in the design matrix. Then a T statistic associated with the regression parameter $\beta$ was formed and thresholded at \(p<0.001\) (one-sided), uncorrected for multiple comparisons. %

As we know the ground-truth activated ROI, we can determine which detected activations are the true/false positives and negatives. The region of interest being small, we have a strongly imbalanced dataset and used the Precision/Recall curve instead of the classical receiver operator characteristic (ROC) curve to accurately compute for each scenario both the area-under-curve (AUC) and balanced accuracy (BACC) scores at \(p<0.001\).
Additionally, we also measured image quality metrics (PSNR and SSIM) as well as temporal SNR (tSNR), to assess the quality of the reconstruction in space and time.

\section{Results}
\label{sec:scenario-results}

\subsection{Scenario S1}

\begin{figure}[hbtp]
  \begin{subfigure}[t]{1.0\linewidth}
    \centering
    \includegraphics[width=0.9\textwidth]{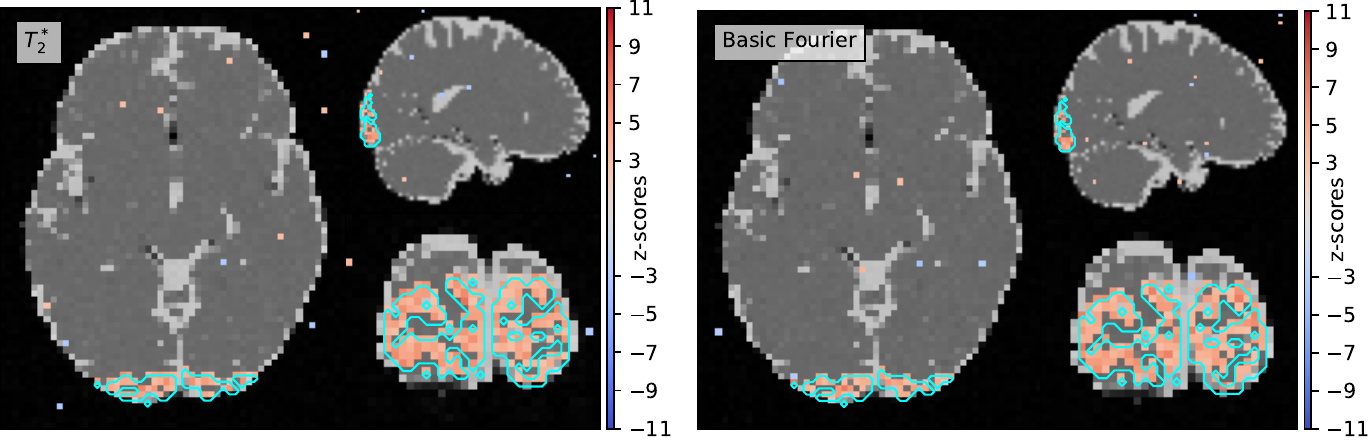}
    \caption{Reconstructed fMRI volume for Scenario S1 for \Ts{} model \eqref{eq:fourier_ext_full} (left) and Basic Fourier model~\eqref{eq:fourier-not2} (right). The ground-truth ROI where activation lies is outlined in cyan. Detected activations surviving at \(p<0.001\), uncorrected, thresholding are overlaid using a colorjet map.}
    \label{fig:scenario1-image}
  \end{subfigure}\vspace*{1em}
    \begin{subfigure}[t]{0.6\linewidth}
        \centering
        \includegraphics[height=0.55\linewidth, valign=t]{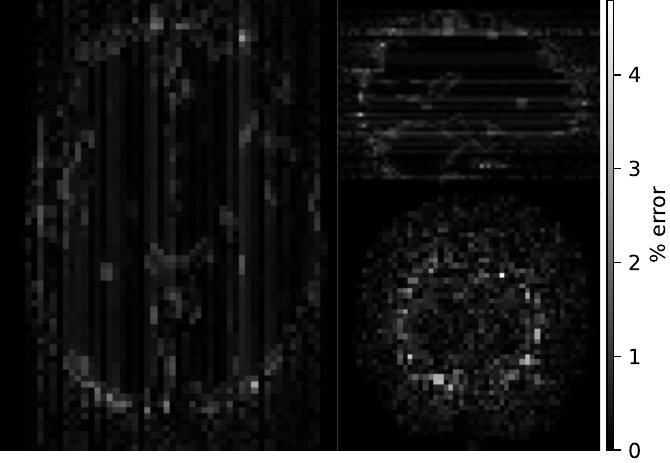}
        \captionsetup{width=.7\linewidth}
        \caption{Error between the \Ts{} and Basic Fourier models, using the first reconstructed volume of each time series. The error has been rescaled by the maximum intensity of the basic Fourier model.}
        \label{fig:scenario1-relerr}
      \end{subfigure}\hfill%
    \begin{subfigure}[t]{0.4\linewidth}
        \centering
        \includegraphics[width=0.9\linewidth, valign=t]{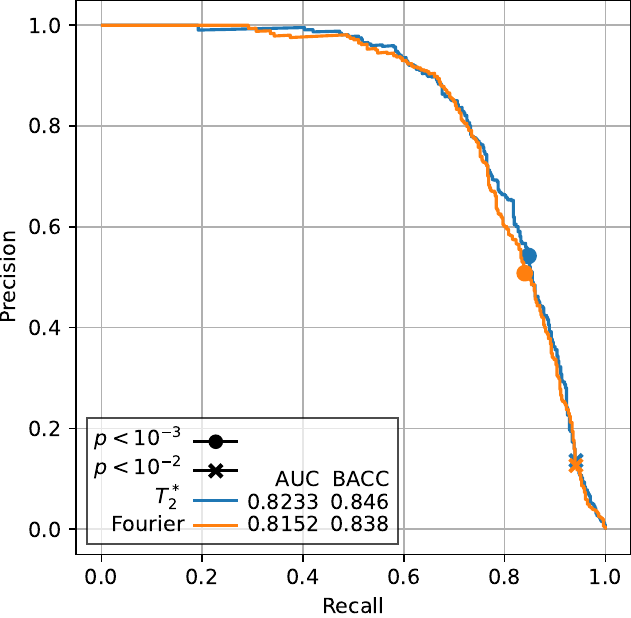}
        \caption{Precision/Recall curves for S1 for the two models.}
        \label{fig:scenario1-pr}
      \end{subfigure}
\caption{Comparison of the two acquisition models~(basic Fourier and \Ts{} effects) for scenario S1.}
\end{figure}

In Scenario 1, the k-space was fully sampled using EPI planes and reconstructed using the inverse FFT. The low spatial and temporal resolution results in high SNR and good image quality, as shown in \Cref{fig:scenario1-image}.
\Ts{} relaxation only introduces negligible artifacts (5\% of maximal error, see~\Cref{fig:scenario1-relerr}), and does not affect statistical performance compared to the basic Fourier model, as illustrated in \Cref{fig:scenario1-pr}. In general, this scenario validates the ability of \SNAKE{} to handle end-to-end fMRI simulation and reconstruction.

\subsection{Scenario S2}
Scenario S2 focuses on the joint effect of optimizing acquisition and reconstruction strategies for acceleration purposes and, therefore, demonstrates the versatility of \SNAKE{}.
As described in \Cref{sec:scenario2-desc} two acquisition strategies (static vs. dynamic spiral stack) compete and three variations of reconstruction methods~(cold vs warm vs refined initialization for each volume).

\begin{figure}[hbtp]
  \centering
  \begin{subfigure}[t]{0.9\linewidth}
    \caption{\Ts{} model \label{subfig:T2smodel_SoS}}
    \includegraphics[width=\textwidth]{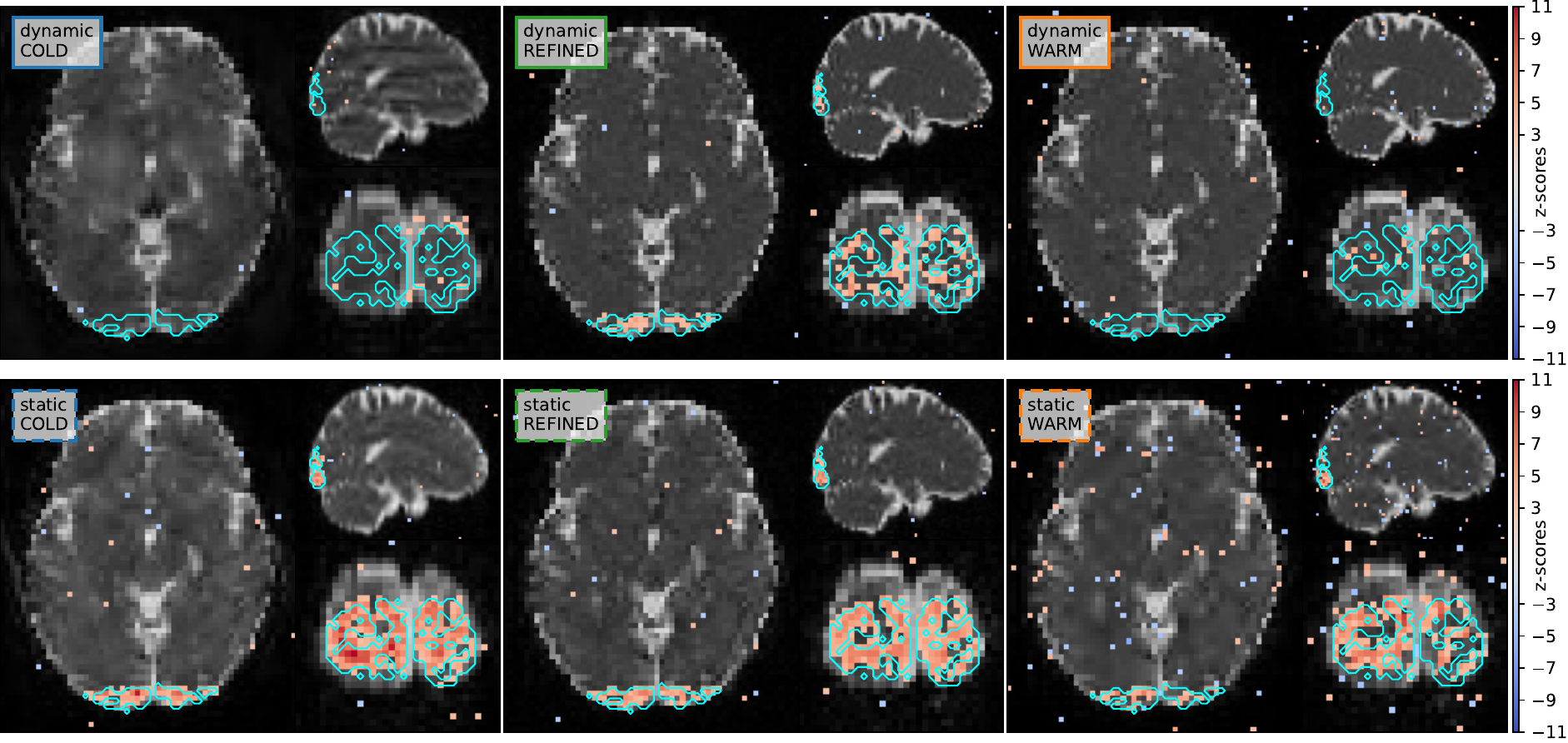}
  \end{subfigure}~\\[.5cm]
  \begin{subfigure}[t]{0.9\linewidth}
    \caption{Basic Fourier model \label{subfig:Fmodel_SoS}}
    \includegraphics[width=\textwidth]{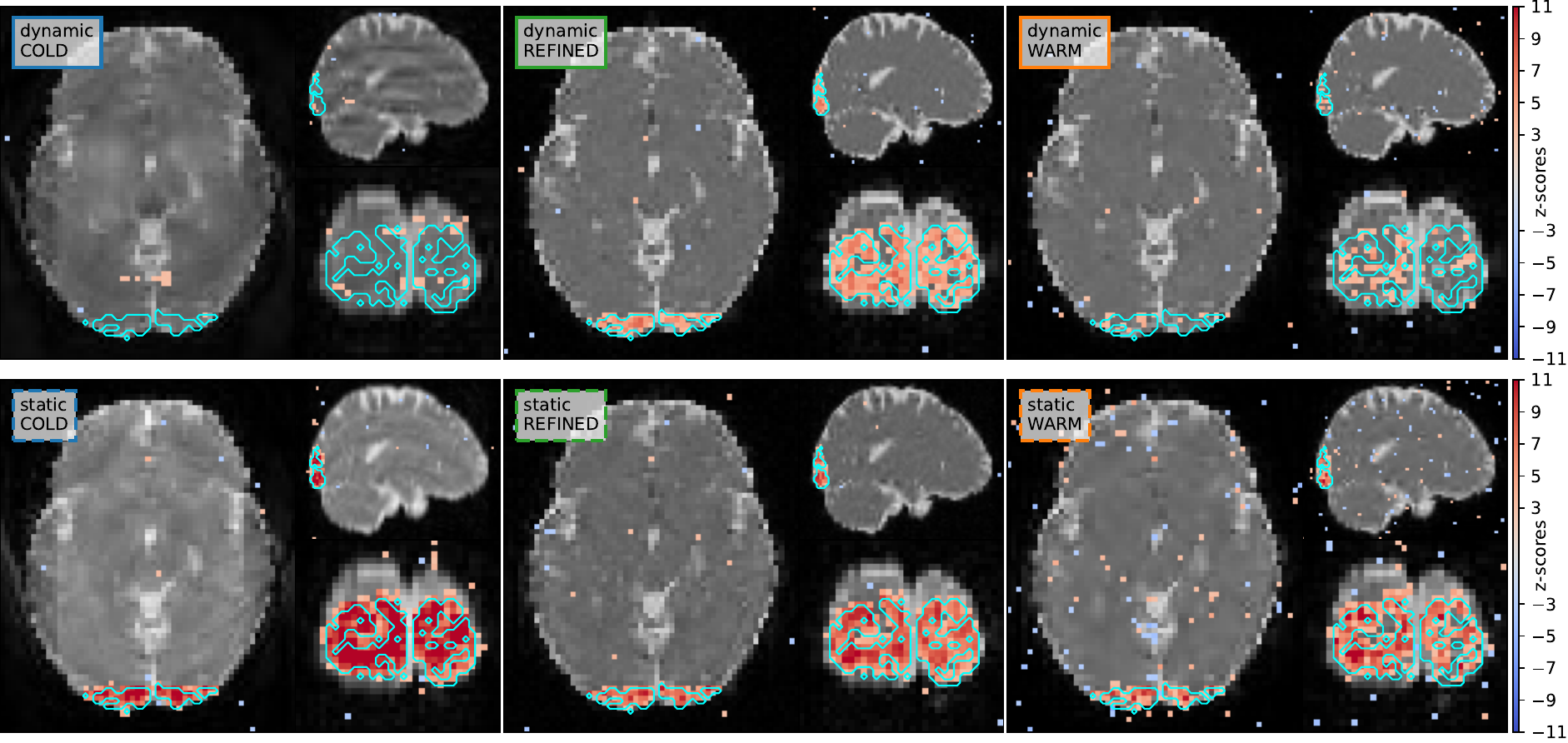}
  \end{subfigure}
  \caption{\label{fig:scenario2-activations}Activations maps for scenario S2. Top: \Ts{} relaxation is taken into account. Bottom: Basic Fourier model, no \Ts{} relaxation.
  Colored frames within each insert follow the convention adopted in the tables reported in \Cref{fig:scenario2-quant}. Detected activations surviving at \(p<0.001\), uncorrected, thresholding are overlaid using a colorjet map.}
\end{figure}

\begin{figure}[hbt]
  \centering
  \begin{subtable}[c]{\linewidth}
    \caption{\Ts{} model \label{tab:scenario2-quant_T2s}}
    \adjustbox{valign=t}{\begingroup\footnotesize
\begin{tblr}{width=\textwidth,
cllrrrrrrr, 
vline{1,Z}={0.5pt},
vline{4,6,8,9}={0.5pt},
hline{1,Z}={0.5pt},
hline{2}={1-7}{0.5pt},
hline{3,6}={0.5pt},
colsep=4pt
}
    \SetCell[c=3,r=1]{c} Setup &        &        & \SetCell[c=2,r=1]{c} SSIM &                 & \SetCell[c=2,r=1]{c} PSNR &                  & \SetCell[c=1,r=2]{c} tSNR &\SetCell[c=1,r=2]{c} AUC &  \SetCell[c=1,r=2]{c} BACC \\
    Label            &  Acq.           &  Recon. & \SetCell{r}First          & \SetCell{r}Last & \SetCell{r}First          & \SetCell{r}Last  &                            &                         &                             \\
        {\color[HTML]{1f77b4}\tikz\draw[solid, line width=0.5mm](0,0) ++(0,0.1) -- ++(0.7,0.0);} & dynamic & COLD  & {0.64}  & {0.51}  & {19.94}  & {19.80}  & {6.21}  & {0.01}  & {0.55}  \\
        {\color[HTML]{2ca02c}\tikz\draw[solid, line width=0.5mm](0,0) ++(0,0.1) -- ++(0.7,0.0);} & dynamic & REFINED  & \textbf{0.67}  & {0.69}  & \textbf{21.08}  & \underline{21.03}  & {23.64}  & {0.34}  & {0.80}  \\
        {\color[HTML]{ff7f0e}\tikz\draw[solid, line width=0.5mm](0,0) ++(0,0.1) -- ++(0.7,0.0);} & dynamic & WARM  & {0.64}  & \textbf{0.79}  & {19.94}  & \textbf{21.06}  & {12.95}  & {0.01}  & {0.58}  \\
        {\color[HTML]{1f77b4}\tikz\draw[dashed, line width=0.5mm](0,0) ++(0,0.1) -- ++(0.7,0.0);} & static & COLD  & {0.64}  & {0.64}  & {19.91}  & {19.91}  & \textbf{43.08}  & \underline{0.70}  & \textbf{0.96}  \\
        {\color[HTML]{2ca02c}\tikz\draw[dashed, line width=0.5mm](0,0) ++(0,0.1) -- ++(0.7,0.0);} & static & REFINED  & {0.60}  & {0.60}  & \underline{20.46}  & {20.46}  & \underline{37.99}  & \textbf{0.75}  & \underline{0.95}  \\
        {\color[HTML]{ff7f0e}\tikz\draw[dashed, line width=0.5mm](0,0) ++(0,0.1) -- ++(0.7,0.0);} & static & WARM  & \underline{0.64}  & \underline{0.74}  & {19.91}  & {20.26}  & {20.18}  & {0.35}  & {0.84}  \\
    \end{tblr}\endgroup
}%
    \hfill\includegraphics[width=0.3\textwidth, valign=t]{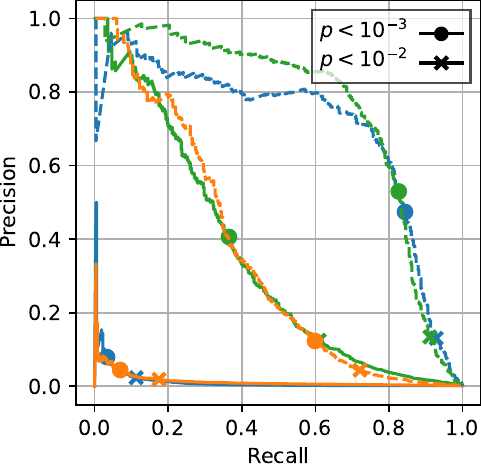}
  \end{subtable}
  \begin{subtable}[c]{\linewidth}
    \caption{Basic Fourier model \label{tab:scenario2-quant_F}}
    \adjustbox{valign=t}{\begingroup\footnotesize
\begin{tblr}{width=\textwidth,
cllrrrrrrr, 
vline{1,Z}={0.5pt},
vline{4,6,8,9}={0.5pt},
hline{1,Z}={0.5pt},
hline{2}={1-7}{0.5pt},
hline{3,6}={0.5pt},
colsep=4pt
}
    \SetCell[c=3,r=1]{c} Setup &        &        & \SetCell[c=2,r=1]{c} SSIM &                 & \SetCell[c=2,r=1]{c} PSNR &                  & \SetCell[c=1,r=2]{c} tSNR &\SetCell[c=1,r=2]{c} AUC &  \SetCell[c=1,r=2]{c} BACC \\
    Label            &  Acq.           &  Recon. & \SetCell{r}First          & \SetCell{r}Last & \SetCell{r}First          & \SetCell{r}Last  &                            &                         &                             \\
        {\color[HTML]{1f77b4}\tikz\draw[solid, line width=0.5mm](0,0) ++(0,0.1) -- ++(0.7,0.0);} & dynamic & COLD  & {0.62}  & {0.50}  & {22.92}  & {22.85}  & {9.42}  & {0.06}  & {0.60}  \\
        {\color[HTML]{2ca02c}\tikz\draw[solid, line width=0.5mm](0,0) ++(0,0.1) -- ++(0.7,0.0);} & dynamic & REFINED  & \textbf{0.72}  & {0.74}  & \textbf{27.47}  & \textbf{27.20}  & {33.30}  & {0.78}  & \underline{0.97}  \\
        {\color[HTML]{ff7f0e}\tikz\draw[solid, line width=0.5mm](0,0) ++(0,0.1) -- ++(0.7,0.0);} & dynamic & WARM  & {0.62}  & \textbf{0.85}  & {22.92}  & \underline{27.19}  & {20.00}  & {0.16}  & {0.73}  \\
        {\color[HTML]{1f77b4}\tikz\draw[dashed, line width=0.5mm](0,0) ++(0,0.1) -- ++(0.7,0.0);} & static & COLD  & {0.62}  & {0.62}  & {22.95}  & {22.96}  & \textbf{71.01}  & \underline{0.79}  & {0.97}  \\
        {\color[HTML]{2ca02c}\tikz\draw[dashed, line width=0.5mm](0,0) ++(0,0.1) -- ++(0.7,0.0);} & static & REFINED  & \underline{0.65}  & {0.66}  & \underline{25.13}  & {25.14}  & \underline{49.36}  & \textbf{0.88}  & \textbf{0.99}  \\
        {\color[HTML]{ff7f0e}\tikz\draw[dashed, line width=0.5mm](0,0) ++(0,0.1) -- ++(0.7,0.0);} & static & WARM  & {0.62}  & \underline{0.78}  & {22.95}  & {24.51}  & {28.77}  & {0.63}  & {0.94}  \\
    \end{tblr}\endgroup
}%
    \hfill\includegraphics[width=0.3\textwidth, valign=t]{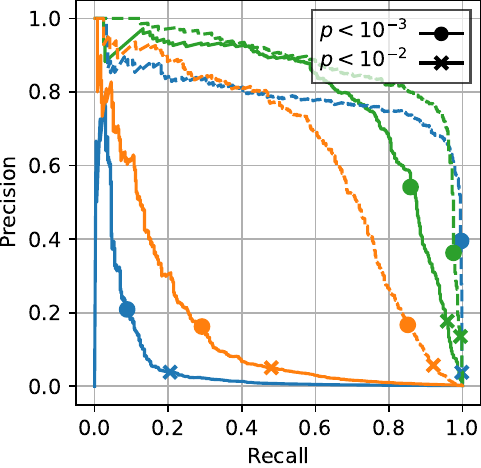}
  \end{subtable}
  \caption{
  \label{fig:scenario2-quant}
  Quantitative metrics summarizing image quality~(SSIM/PSNR), signal quality~(tSNR) and statistical performances~(AUC/BACC) for scenario S2. Top: \Ts{} relaxation is taken into account. Bottom: Basic Fourier model. TSNR values are averaged over the occipital ROI. In the tables, First and Last refer to the first and last frames in the fMRI volumes sequences.}
\end{figure}

The image quality and statistical maps are shown in \Cref{fig:scenario2-activations} and the extensive quantitative statistical analysis is reported in
\Cref{fig:scenario2-quant}. Based on these results we can make the following claims.

\subsubsection{\Ts{} relaxation must be taken into account for the long readout in non-Cartesian acquisitions}

The \Ts{} relaxation model must not be neglected in the simulation stage when considering non-Cartesian readouts instead of 3D EPI for comparable \TE{} and observation time $T_{obs}$, as its impact on both image quality and statistical performance is significant. Notably, the comparison of results shown in \Cref{tab:scenario2-quant_T2s} and \Cref{tab:scenario2-quant_F} depicts a significant drop in SSIM, PSNR and tSNR scores for the \Ts{} relaxation model due to blurring and contrast loss. This decrease in image quality happens because the reference MR image is the ideal phantom with the contrast at \TE{}, which does not suffer from the \Ts{} relaxation.
Similarly, we observe a decrease in AUC/BACC for the \Ts{} relaxation model, which is due to a systematic lower precision~(higher false positive rate). This loss in precision is the result of lower z scores for that model compared to the basic Fourier model, as shown when contrasting \Cref{subfig:T2smodel_SoS} and \Cref{subfig:Fmodel_SoS}.

\emph{Hereafter, we only discuss the \Ts{} relaxation model for this scenario, as it is the closest to real fMRI acquisition.}

\subsubsection{The refined initialization is key in dynamic acquisition for improved performances}

When comparing image reconstruction strategies, we observe in \Cref{fig:scenario2-quant} that the warm-start mechanism is beneficial to improve the image quality between the first and last frames for both models.
The gain in both SSIM and PSNR scores is actually large when this strategy is turned on.
However, activating only the first warm start strategy does not yield improved statistical performance.
Instead, using the \emph{dynamic refined} approach in the reconstruction stage adds value to image quality and statistical performance, notably for the dynamic acquisition setup.
In this specific scenario, this reconstruction strategy allows each frame to actually bring new information from complementary under-sampled k-space data and thus maintain good performances while increasing temporal resolution.

Moreover, the use of time-varying under-sampling patterns introduces temporal incoherence across frames and thus introduce different aliasing artifacts for each frame, which can be detrimental to image quality and statistical performances~(see for instance the large gaps in the different scores between the \emph{dynamic cold} and \emph{refined} strategies in \Cref{fig:scenario2-quant}).

\subsubsection{Image quality is not a right proxy for good statistical performances}

The static acquisition strategy provides the best statistical results, as it allows us to detect evoked brain activity in the targeted ROI with the highest precision/recall and tSNR scores~(cf.
\Cref{tab:scenario2-quant_T2s}).
Here, the tSNR metric correlates well with AUC/BACC scores as the noise is purely thermal~(no physiological noise).
However, image quality is degraded in these cases; this is notably visible in \Cref{fig:scenario2-activations} when comparing the results \emph{static cold} to \emph{dynamic refined}.
Strong aliasing and blurring artifacts actually corrupt reconstructed MR images in the static regime.
Additionally, consistency of image quality (even in the degraded case) over the whole fMRI run also matters, as warm-start reconstruction (where image quality improves over time) shows poor statistical performances.

Finding the best trade-off between image quality and statistical performance can be achieved using the refined mechanism in terms of initialization strategy, as it provides the best image quality throughout the sequence of fMRI volumes while retaining good statistical performance.
This is particularly noticeable in the dynamic acquisition setting: The AUC/BACC scores associated with the refined strategy~(solid green curve) are higher than those associated with warm/cold strategies~(solid blue and orange curves); see \Cref{tab:scenario2-quant_T2s} for the \Ts{} model.
More strikingly, this also happens in the static acquisition setting.
A significant gain in SSIM/PSNR is observed with the refined strategy, and still the AUC/BACC scores are comparable to those measured using the cold initialization.
When comparing the green and blue Precision / Recall dashed curves in \Cref{tab:scenario2-quant_T2s}, one can even observe that the green bullet is on top of the blue one, indicating a slight boost in precision~(statistical sensitivity) at a given specificity level~(same value along the y axis).
Finally, all these observations that were established on the \Ts{} model remain valid in the simplified Fourier model~(cf.~\Cref{tab:scenario2-quant_F}), which enforces these statements.

However, dynamic acquisition strategies have shown some great potential for increasing temporal resolution while maintaining good image quality. Preserving fine details in fMRI images at the output of the reconstruction pipeline matters for correct preprocessing of fMRI volumes and reliable statistical analysis (registration to a template, extraction of the cortical surface, etc.).

\subsection{Scenario S3}
Last, we present the results obtained for Scenario S3 (1mm isotropic resolution, \(\TR_{vol}\) = 2.4 s) as a proof of concept of the scalability of \SNAKE{} to high spatial resolution non-Cartesian fMRI.

\begin{figure}[h!]
  \centering
  \begin{subfigure}[t]{0.7\linewidth}
    \centering
    \caption{Activation maps. Top: Fourier model. Bottom: \Ts{} model.\label{fig:S3_activation_maps}}
    \includegraphics[width=0.95\textwidth,valign=t]{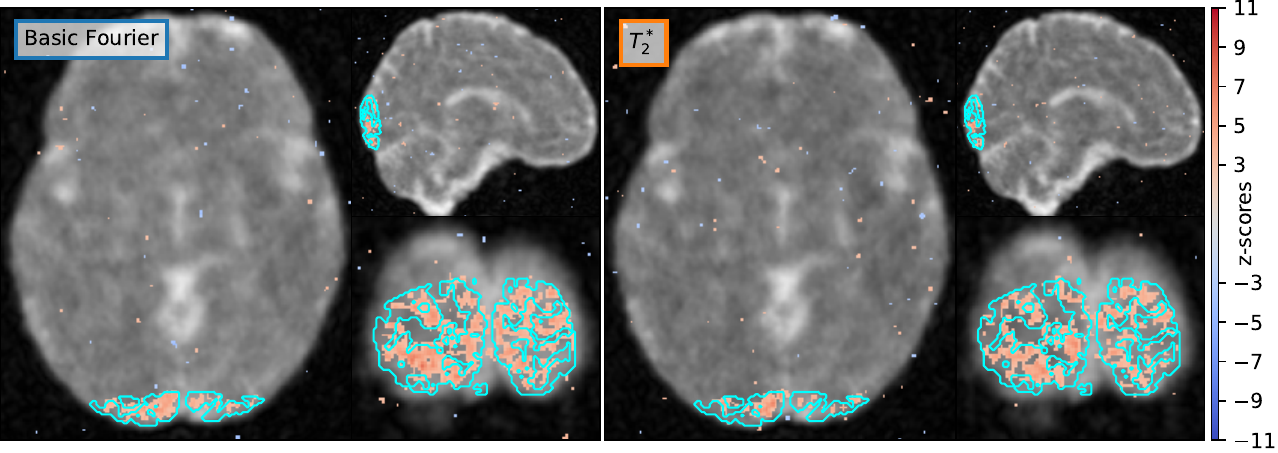}
  \end{subfigure}\hfill
  \begin{subfigure}[t]{0.3\linewidth}
    \caption{Precision Recall Curve.}
    \includegraphics[width=0.9\textwidth, valign=t]{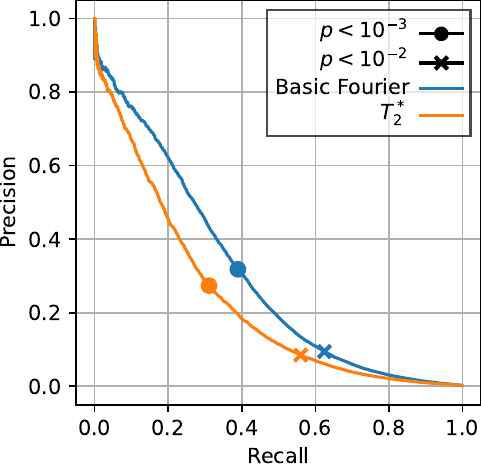}
  \end{subfigure}\\
  \begin{subtable}[t]{0.6\linewidth}
    \centering
    \caption{Quantitative results for Scenario 3.\label{tab:S3_metrics}}
    \begingroup\footnotesize
\begin{tblr}{width=\textwidth,
clrrrrrrr,
vline{1,Z}={0.5pt},
vline{3,5,7,8}={0.5pt},
hline{1,Z}={0.5pt},
hline{2}={1-7}{0.5pt},
hline{3}={0.5pt},
colsep=4pt
}
    \SetCell[c=2,r=1]{c} Setup &  & \SetCell[c=2,r=1]{c} SSIM &     & \SetCell[c=2,r=1]{c} PSNR &   &\SetCell[c=1,r=2]{c} tSNR  & \SetCell[c=1,r=2]{c} AUC &  \SetCell[c=1,r=2]{c} BACC \\
    Label            &  Acq.   & \SetCell{r}First & \SetCell{r}Last & \SetCell{r}First & \SetCell{r}Last &                      &                          &                             \\
        {\color[HTML]{1f77b4}\tikz\draw[solid, line width=0.5mm](0,0) ++(0,0.1) -- ++(0.7,0.0);} & Basic Fourier  & \textbf{0.306}  & \textbf{0.309}  & \textbf{22.204}  & \textbf{22.220}  & \textbf{36.224}  & \textbf{0.298}  & \textbf{0.806}  \\
        {\color[HTML]{ff7f0e}\tikz\draw[solid, line width=0.5mm](0,0) ++(0,0.1) -- ++(0.7,0.0);} & $T_2^*$  & \underline{0.305}  & \underline{0.307}  & \underline{22.100}  & \underline{22.115}  & \underline{34.894}  & \underline{0.229}  & \underline{0.774}  \\
    \end{tblr}\endgroup

  \end{subtable}

\caption{Results for scenario S3. Top left/center: First fMRI reconstructed volume using the standard Fourier~(left) and \Ts{}~(center) models. Sequential cold-start CS reconstruction was performed frame-wise. Detected activations surviving at $p<0.001$ thresholding are overlaid using a colorjet map.
Top right: Precision/Recall curves for the two models. Accurate modeling of realistic \Ts{} decay slightly impacts statistical performances. Bottom: Quantitative metrics summarizing image quality~(SSIM/PSNR), signal quality~(tSNR) and statistical performances~(AUC/BACC).\label{fig:scenario3}}
\end{figure} 

With full GPU acceleration, sequential cold start reconstruction took 4min4sec per frame to converge, compared to 15-20min in the previous implementation~\citep{amor2024non}, thus requiring 8h30min for a complete fMRI run~(125 frames).
Both models (with and without \Ts relaxation) show similar and relatively poor image quality due to blurring, as shown in \Cref{fig:S3_activation_maps}. Additionally, the statistical performances are slightly lower~(loss in sensitivity and specificity) when the \Ts relaxation effect is considered in the data simulation process, as illustrated in \Cref{tab:S3_metrics}.
\section{Discussion}

\label{sec:discussion}

\subsection{Effect of \Ts{} relaxation}

All scenarios have roughly the same \Ts{}, \TE{} and \(T_{obs}\), but the \Ts{} relaxation effect depends on the acquisition strategy and the k-space trajectory used.
In Scenario 1 (EPI trajectory) no noticeable changes are visible, as the k-space is fully sampled and nearby k-space points are acquired at close times.
In Scenario 2, the spiral trajectory allows for the acquisition of two neighboring points in k-space at both temporal extremities of the spiral.
This results in a change of imaging contrast between two neighboring measurements in the k-space, thus inducing a loss of contrast in the reconstructed fMRI images.
This effect can be alleviated using temporally smooth trajectories such as cones or MORE-SPARKLING ~\citep{giliyar_radhakrishna_improving_2023} readouts.
Statistical performance is also affected as lower z scores were retrieved for the \Ts{} model compared to the basic Fourier model, as shown in \Cref{fig:scenario2-activations}.
However, for well-chosen reconstruction scenarios~(e.g.
static cold/refined), evoked activity is still well detected in the occipital ROI, and the precision-recall at \(p<0.001\) remains similar.
Similar findings were observed in the case of dynamic acquisition with refined reconstruction.

In general, we recommend keeping the modeling of the relaxation effects \Ts{} in the simulation despite the higher computing cost, as they can still influence the extraction of z scores as a function of the readout of k-space for given \TE{} and \(T_{obs}\).

\subsection{Setting the noise level and tSNR}

Concerning the choice of \(SNR_{i}\), we reported in \Cref{sec:appendic_tSNR} the tSNR maps for each scenario. In S1 we measured a tSNR of 40.5 for the basic Fourier and $\Ts{}$ models in the occipital ROI. The tSNR increased to 120 in the CSF as shown in \Cref{fig:scenario1-tsnr}. This is in line with what we could expect under experimental conditions. TSNR values for S2 are reported in \Cref{fig:scenario2-quant}. As the \Ts{} relaxation is significant in this scenario, we can see a drop of tSNR, which goes along with lower statistical performances. However, tSNR is not an oracle for statistical performance, as outlined in \citep{jamil_temporal_2021}. Finally, in S3, we get a tSNR of 36 in the occipital ROI. By decreasing \SNRi{} at higher resolution from 1000 at 3mm iso to 30 at 1mm iso, we obtained similar tSNR ranges in the three scenarios. Note that for S3, similar values have been empirically observed in \citep{amor_prospects_2022}.

\subsection{Comparison between scenarios}

The three scenarios presented in \Cref{sec:scenario-desc} and \ref{sec:scenario-results} provide complementary views on the comparison of acquisition and reconstruction strategies.

Scenario S1 covers a low spatial and temporal resolution setting, uses standard Cartesian acquisition and basic reconstruction methods, and shows the potential of \SNAKE{} to provide a basic reference for detecting evoked brain activity at a given input SNR.
Beyond providing a simple and fast validation case for the simulator, this scenario could be used to compare the statistical performances of different competing experimental paradigms regarding the number and duration of stimuli, block vs. event-related designs, for a given scan time budget and under various artifacts.

Scenario S2 explores a low spatial, high temporal resolution setup and provides a large panorama of possible acquisition and reconstruction strategies.
It shows the potential of \SNAKE{} to produce a benchmark of competing techniques both at acquisition~(static vs dynamic stack of spirals) and image reconstruction stages, particularly in the CS setting.
This benchmark could be extended in the near future by using more complex reconstruction methods such as ow-rank~+~sparse methods or even deep learning approaches.

Interestingly, from its current implementation, we observed that the best image quality is not always associated with the highest statistical scores, and that the refined initialization strategy is key in dynamic acquisition for improved statistical performances.

Finally, Scenario S3 provides a high spatial and low temporal resolution setup, reaching the limits of current fMRI acquisition strategies for whole brain coverage.
Unlike S2, the SPARKLING under-sampling pattern does not show aliasing artifacts, even at high spatial resolution, however, at the cost of some blurring.
More generally, \SNAKE{} offers new possibilities to optimize further 3D and even 4D under-sampling patterns.

In general, these three scenarios should be considered as an upper bound in terms of image quality and statistical performances for prospective validation on real fMRI experiments, as actual fMRI data acquisition and image reconstruction face additional issues such as imperfection in coil sensitivity estimation, presence of motion artifacts, off-resonance effects due to static and dynamic $B_0$ inhomogeneities.

\subsection{Limits to the study}

Our work is well defined by the hypothesis formulated in \Cref{sec:summary-hypotheses}, and their practical application in \Cref{sec:scenario-desc}.
In particular, the scenarios presented in this paper omit three major sources of noise in the fMRI data: head movement, physiological effects, and inhomogeneities of $B_{0}$, since we focus on the core capabilities of \SNAKE{}.
Similarly, we only used the three main tissue classes (WM, GM, CSF).
This allowed us to speed up the acquisition.
However this choice was also the consequence of missing information in the literature about relaxation parameter values for other tissue classes at 7T.
Moreover, the Cartesian scenario restricts itself to fully sampled EPI and lacks any acceleration setup such as GRAPPA or CAIPI~\citep{breuer_controlled_2005, griswold_generalized_2002}. Future work will address these aspects and explore more challenging reconstruction setups. However, \SNAKE{} provides neuroscientists with an upper limit on the statistical impact and performance of many combinations of acquisition and reconstruction strategies, which offers an easy exploration of new methodologies.

\subsection{Extending the simulator}

As \SNAKE{} is an open source software, external contributors from the fMRI community are welcome to participate in its extension to help refine the forward modeling to handle multiple sources of artifacts that contaminate actual fMRI data.
In addition to head motion and off-resonance (of which we are adding some preliminary support),
one may think of modeling of temporal aliasing artifacts due to physiological rhythms~(heart beat or breathing rate) that are not sampled fast enough.
Adequate fMRI acquisition and reconstruction methods could be studied with \SNAKE{} to mitigate these additional sources of disturbance.

Similarly, more complex brain activation patterns spread over multiple ROIs (for using functional atlases from  \citet{yeo_organization_2011}) could be designed with spatial variations in the HRF shape, following the seminal work of PyHRF~\citep{vincent_flexible_2014}, or using meta-analyses and Python tools like neurosynth\footnote{\url{https://neurosynth.org/}.}~\citep{yarkoni2011large} or Neuroquery\footnote{\url{https://neuroquery.org/query?text=number+computation}.}\citep{dockes2020neuroquery} to define well-located activation peaks for given cognitive paradigms and tasks.
So far, \SNAKE{} has focused on brain mapping tasks in task-related fMRI. However, it could address the simulation and analysis of resting-state fMRI and be tuned to optimize the retrieval of resting-state networks from synthetic rs-fMRI data sets.
In that regard, coupling with other fMRI simulators, such as Virtual Brain~\citep{schirner_brain_2022}, could be instrumental in generating realistic rs-fMRI data as input reference data to \SNAKE. As \SNAKE{} is already interfaced with the nilearn package for statistical analysis and because the latter allows functional connectivity analysis, this extension to rs-fMRI data analysis could be quite straightforward.

However, the complexity of the simulation should be balanced with the need for the explicability of the results.
On the one hand, adding layers (and their potential heavy parameterization) to the simulation will also induce a potential loss in explicability for the effect on downstream applications.
On the other hand, the modularity of \SNAKE{} allows the ablation study to enable / disable any aspect of the simulation that is to be analyzed which has the most significant impact.

\subsection{Exploring the effect of tuning acquisition and reconstruction together}

The first results obtained with \SNAKE{} show the criticality of matching and tuning the experimental design, acquisition, and reconstruction strategies to obtain the best quality in the subsequent analysis.
More complex acquisition and reconstruction methods could leverage the temporal redundancy in k-space data with global \emph{a priori} such as low rank + sparse~\citep{petrov_improving_2017,graedel_ultrahigh_2022}. The \emph{refined} strategy introduced in Scenario S2 shows some great potential for both static and dynamic acquisition schemes, and will be at the core of future development.

To our knowledge, as summarized in \Cref{tab:summary-sim}, \SNAKE{} is the only open source fMRI data simulator that can efficiently provide arbitrary fMRI k-space data.
It is also the only one that can be used to benchmark reconstruction methods in an automated manner.

\subsection{More than a fMRI simulation tool}

The modular approach of \SNAKE{} also enables applications other than simple simulations.
First, the combination of the running scheduling and analysis modules provides a reliable benchmark for image reconstruction methods (even for a single anatomical volume).

Second, each handler can also be viewed (and used) as a data augmentation layer for supervised deep learning methods, opening up new opportunities for fMRI image reconstruction. Currently, such models cannot be trained in the supervised setting due to the lack of ground truth~(i.e. non-accelerated) real high resolution fMRI data. Alternative self-supervised approaches based on domain undersampling~\citep{demirel2021improved} are limited in terms of performance, notably in the high-resolution setting.
Hence, \SNAKE{} could be used to train deep neural networks dedicated to fMRI image reconstruction on synthetic fMRI data. These models might be fine-tuned later with transfer learning on real data sets. Alternatively, \SNAKE{} might serve as a data augmentation tool. The recent review by \citet{gopinath_synthetic_2024} outlined the need for efficient synthetic data generation, a task for which \SNAKE{} was precisely designed.

\raggedbottom
\section{Conclusion}

In this paper, we have proposed a new fMRI data simulator, called \SNAKE{}, which is packaged as an open source Python software, offering the ISMRM and OHBM communities the opportunity to advance the field of optimal fMRI data acquisition and image reconstruction at low scanning cost.
More specifically, \SNAKE{} is purposely designed to assess the impact of massively undersampled 3D non-Cartesian readouts aiming to reach unprecedented spatial or temporal resolution while maintaining whole-brain coverage, good image quality, and the ability to detect tiny BOLD effect through statistical guarantees. 
Through its modular design, \SNAKE{} has been thought to remain open to external contributions as there is room to model additional aspects, notably external sources of artifacts~(static and dynamic $B_0$ inhomogeneities, motion, etc.).

\SNAKE{} comes also with an end-to-end and extensible reconstruction and statistical analysis pipeline. It provides the user with tools to reach new frontiers in fMRI data acquisition and image reconstruction strategies and to evaluate multiple competing scenarios from an image quality and statistical analysis viewpoint that cannot easily be ranked in advance. Future work will be dedicated to the interface \SNAKE{} with deep learning for fMRI image reconstruction. %

\section*{Declaration of Competing Interest}
The authors declare that they have no competing financial interest.

\section*{Author Contributions}
  \emph{Conceptualization}:  P.-A. C., P.C. and A.V.\\
  \emph{Software, Visualization, Writing - Original Draft}: P.-A. C.\\
  \emph{Writing - Review \& Editing}: P.C. and  A.V.\\
  \emph{Supervision}: P.C. and A.V.\\
  \emph{Funding Acquisition}: P.C.

\section*{Software ecosystem acknowledgment}
The package \SNAKE{} would not have been possible without the following existing software:

The command line and configuration files use \href{https://github.com/facebookresearch/hydra}{\texttt{hydra}}~\citep{yadan_hydra_2019}.
The numerical computation involves \href{https://github.com/numpy/numpy}{\texttt{numpy}}~\citep{harris_array_2020} and \href{https://github.com/scipy/scipy}{\texttt{scipy}}~\citep{virtanen_scipy_2020}.
Neuroscience-related processing is performed using \texttt{nilearn}~\citep{abraham_machine_2014}and statistical analysis is done with \href{https://github.com/scikit-learn/scikit-learn}{\texttt{scikit-learn}}~\citep{scikit-learn}.

\SNAKE{} was developed concurrently with other software as dependencies. MRI-related work uses \href{https://github.com/paquiteau/brainweb-dl}{\texttt{brainweb-dl}} to access the BrainWeb dataset~\citep{aubert-broche_twenty_2006}, and \href{https://github.com/mind-inria/mri-nufft}{\texttt{mri-nufft}} to generate and acquire non-Cartesian trajectories.  The tracking and management of experiments motivated the development of \href{https://github.com/paquiteau/hydra-callbacks}{\texttt{hydra-callbacks}}.

\section*{Data and code availability}
The \SNAKE{} package, its documentation and the scenarios presented are available at \url{https://github.com/paquiteau/snake-fmri}.
\pagebreak

\appendix
\section{BrainWeb Phantom}
\label{sec:brainweb-phantom}
\begin{figure}[hbtp]
  \centering
  \begin{subfigure}{1.0\linewidth}
    \centering
    \includegraphics[width=\linewidth]{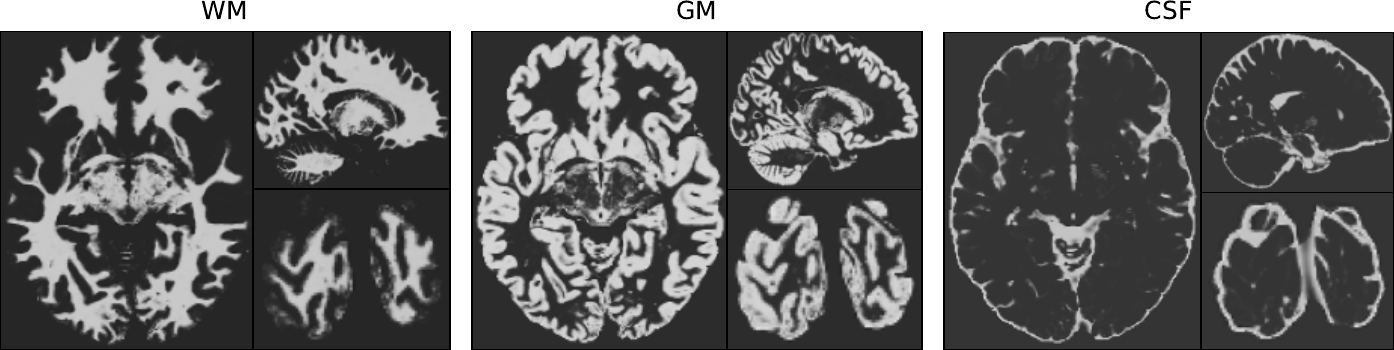}
    \caption{Fuzzy tissues masks at 0.rmm iso resolution.}
  \end{subfigure}\\[1.5em]
  \hfill\begin{subfigure}{0.4\linewidth}
    \centering
    \includegraphics[width=\linewidth]{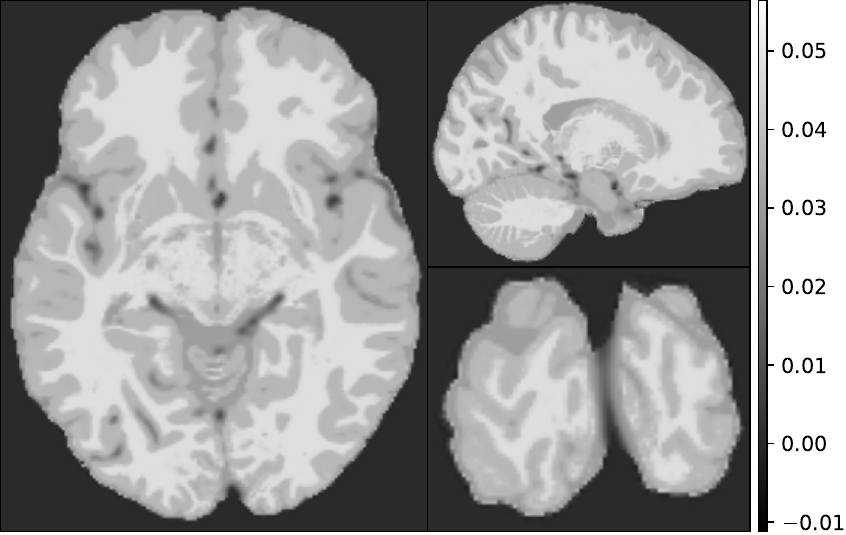}
    \caption{\TE=10ms, \TR=50ms, \FA=45$^{\circ}$.}
  \end{subfigure}\hfill%
  \begin{subfigure}{0.4\linewidth}
    \centering
    \includegraphics[width=\linewidth]{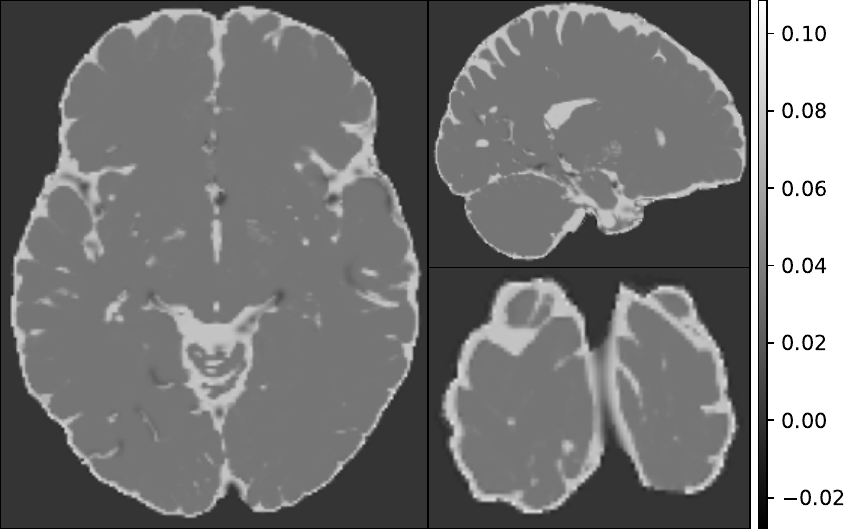}
    \caption{\TE=30ms, \TR=50ms, \FA=12$^{\circ}$.}
  \end{subfigure}\hfill~\\[.5cm]
  \begin{subtable}{0.6\linewidth}
    \centering
    \caption{Tissue Parameters used for the simulation, relaxations times (\(T_{1}\), \(T_{2}\), \(T_{2}^{*}\) are given in ms at 7T), proton density (\(\rho \)) and magnetic susceptibility \(\chi\) are dimensionless.}
      \begin{tblr}{colspec=lrrrrr}
      \toprule
      Tissue & \(T_{1}\)   & \(T_{2}\)   & \(\Ts\)  & \(\rho\)  & \(\chi\)    \\
      \midrule
      WM  & 1200 & 57   & 27   & 0.77 & -9.08  \\
      GM   & 1800 & 49   & 28   & 0.86 & -9.05  \\
      CSF  & 3730 & 1010 & 1010 &1     &  -9.05 \\
      \bottomrule
    \end{tblr}
  \end{subtable}
  \caption{\label{fig:brainweb-phantom} BrainWeb phantom \citep{aubert-broche_twenty_2006} and tissue parameters used for the simulations.}
  
\end{figure}

\section{Estimation of the regularization parameter using SURE.}
\label{sec:app_SURE}

In \Cref{sec:scenario-reconstruction} we explained that the regularization parameter \(\mu_t\) used in \Cref{eq:sequential} was estimated for each frame \(t\). In practice, this is achieved as described in \cref{alg:sure}. As the estimate \(x_t\) becomes cleaner, the threshold automatically decreases. This estimation is very fast to compute because we restrict the estimation to the most detailed wavelet coefficients in the high-frequency subbands. Therefore, the problem dimension is reduced and we can use a line search strategy to find \(\hat{\mu}_t\) without noticeable overhead for image reconstruction. 

\SetKwComment{Comment}{\begingroup \footnotesize }{ \endgroup}
\begin{algorithm}
\DontPrintSemicolon
\SetNoFillComment
\caption{Estimation of \(\mu_t\) for solving \eqref{eq:sequential}\label{alg:sure}}
\KwData{$\bm{x}_t, \bm{\Psi}$}
\KwResult{$\mu_t$}
  $\bm{\alpha}_{H} \gets P^{HHH}(\bm{\Psi}(\bm{x}_t))$ \tcp*{Extract high-details wavelet coeffs}  \;
  $n \gets \textrm{size}(\bm{\alpha}_{H})$ \;
  $\sigma \gets \text{MAD}(\bm{\alpha}_{H}) * 0.675$ \tcp*{Estimate variance using Median Absolute Deviation}\;
  $\bm{\alpha}_{H} \gets \bm{\alpha}_H / \sigma$ \;
  \eIf{$\frac{1}{n}\bm{\alpha}_{H}^T\bm{\alpha}_{H}  < \log_2(n)^{3/2} / \sqrt{n}$}{
  $\mu_{t} \gets \sqrt{2\log_2(n)}$\;
  }{
    $\hat{\mu}_{t} \gets \arg\displaystyle\min_{w \in \bm{\alpha}_{H}} \displaystyle\sum_{i=1}^n \min(\alpha_{H,i}^2, w^2) - 2 * I\{\alpha_{H,i}^2 < w^2\}$ \tcp*{Find a threshold that minimizes the SURE estimator}\;
    $\mu_{t} \gets \min(\hat{\mu}_{t}, \sqrt{2\log_2(n)})$\;
  }
\end{algorithm}

\section{Temporal SNR maps for the three scenarios}\label{sec:appendic_tSNR}

To ensure that the values for \(SNR_{i}\) have been correctly chosen, here we report the tSNR maps for each scenario.
The tSNR in absence of evoked brain activity is defined voxel-wise as:
\begin{align}
\forall v\in \text{FOV}, \quad tSNR(v) &= \operatorname{mean}(x_t(v))/\operatorname{std}(x_t(v)),
\end{align}
where $x_t(v)$ refers to the BOLD fMRI time series at frame $t$ and location $v$.

\begin{figure}[htbp]
\centering
   \begin{subfigure}[b]{0.49\linewidth}
  \centering
    \caption{\label{fig:scenario1-tsnr} tSNR maps for scenario S1. Top: \Ts{} model; Bottom: simple Fourier model.}
  \includegraphics[width=\textwidth]{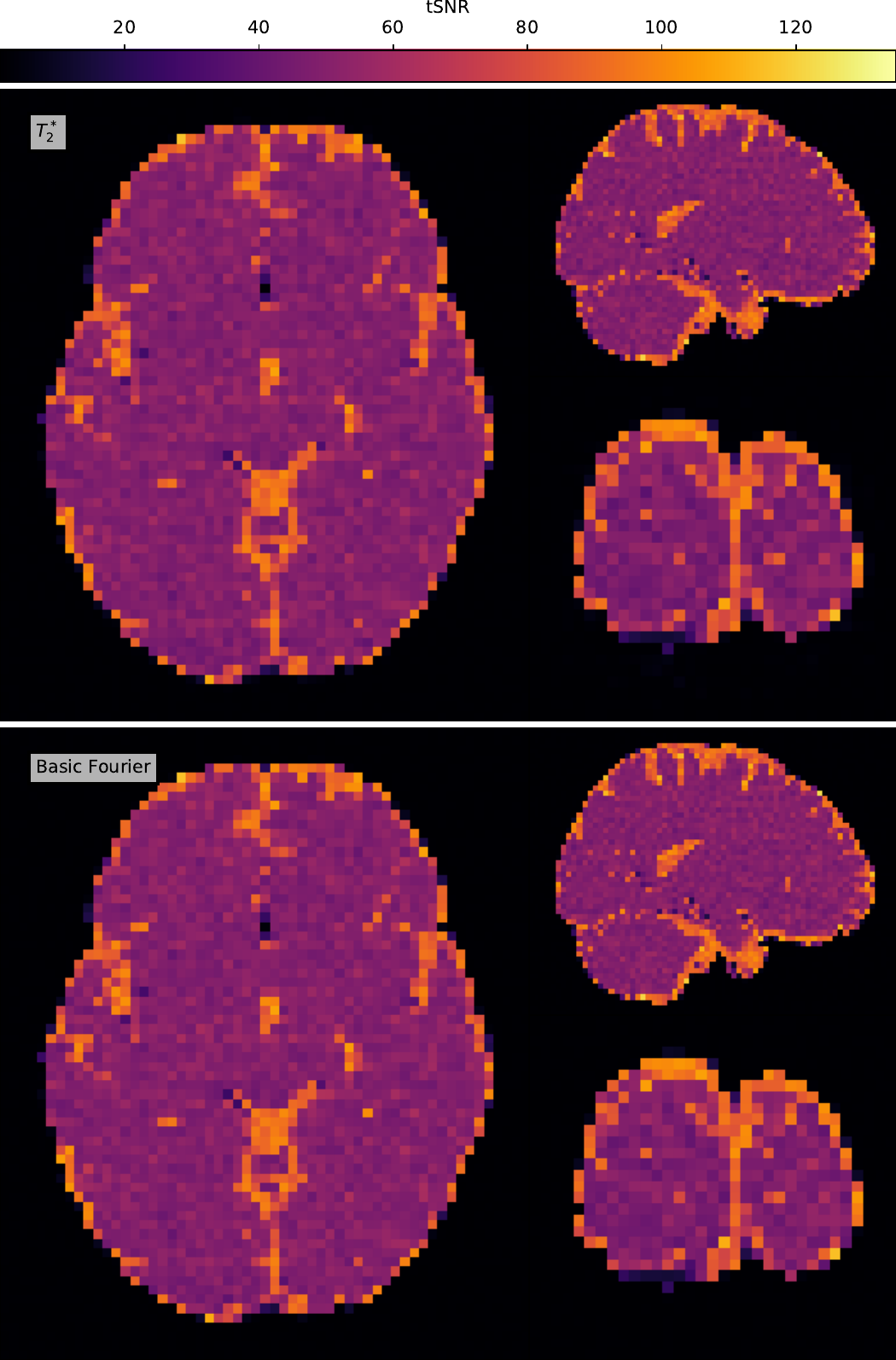}
  \end{subfigure}\hfill
  \begin{subfigure}[b]{0.49\linewidth}
   \centering
   \caption{\label{fig:scenario2-tsnr} tSNR maps for scenario S2 with \Ts{} model. Top/Bottom: Dynamic/Static refined strategy.}
   \includegraphics[width=\textwidth]{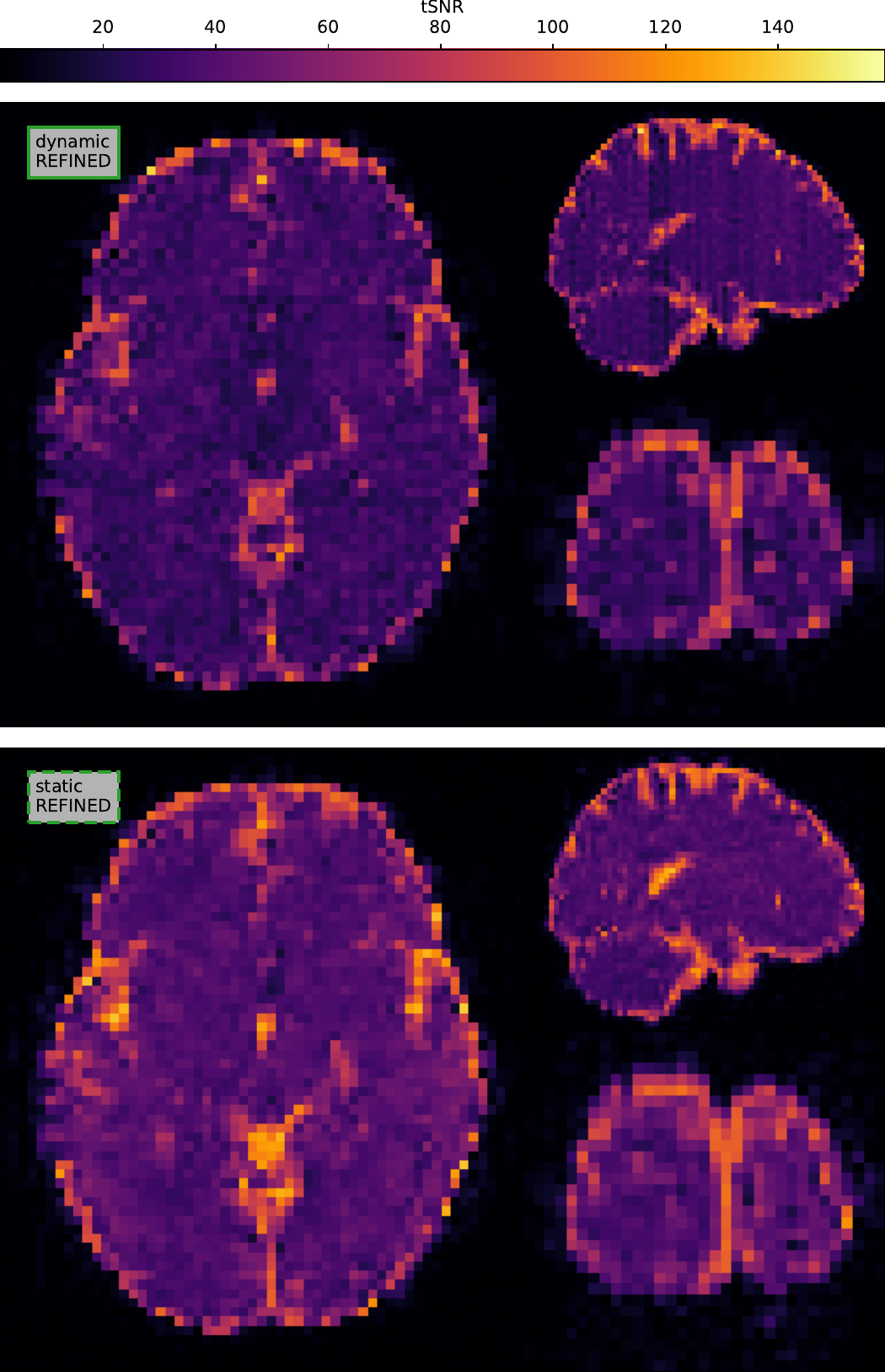}\\
   \end{subfigure}\\[.25cm]
  \begin{subfigure}[b]{0.49\linewidth}
    \centering
       \caption{\label{fig:scenario3-tsnr} tSNR map for scenario S3 with \Ts{} model.}
       \includegraphics[width=\textwidth]{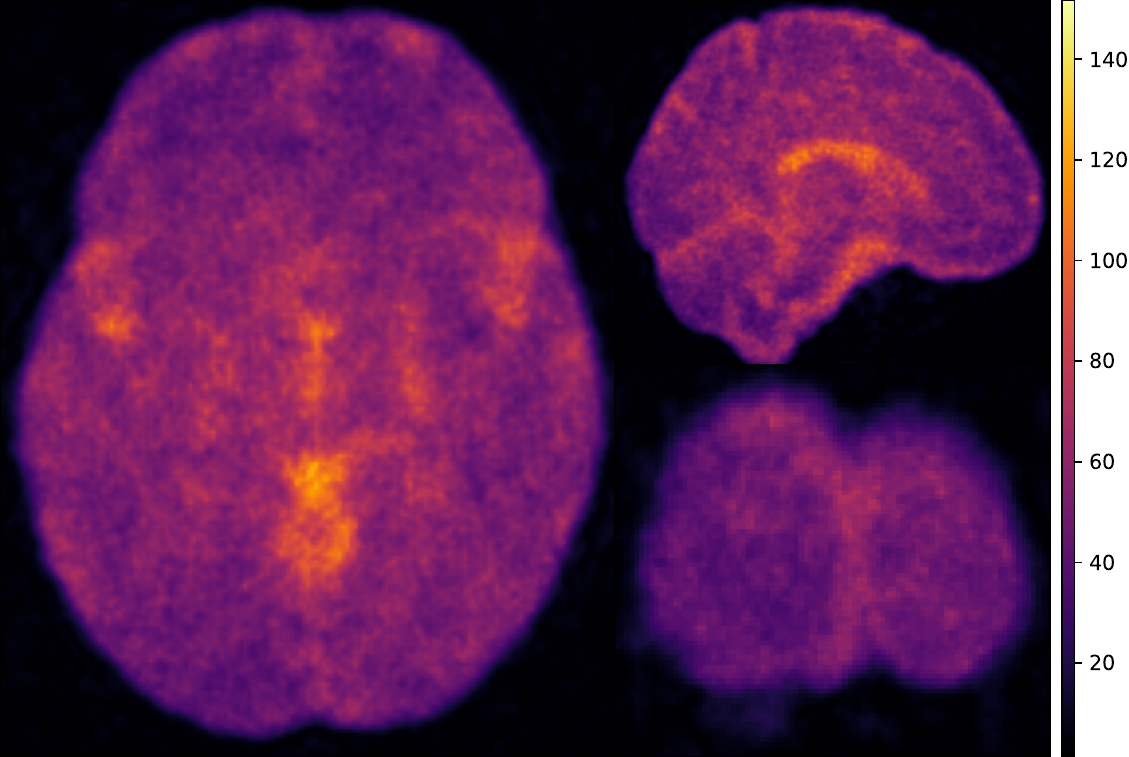}
  \end{subfigure}
  \caption{tSNR maps for the 3 scenarios.}
\end{figure}

Interestingly, in \Cref{fig:scenario1-tsnr} we show that the tSNR is marginally impacted by \Ts{} relaxation effects in Scenario~1 due to the use of Cartesian 3D EPI. In regards to Scenario~2 we already reported that \Ts{} relaxation effects were detrimental to the mean tSNR, mainly due to longer $T_{obs}$.  This is confirmed in \Cref{fig:scenario2-tsnr} with lower tSNR values compared to those shown in \Cref{fig:scenario1-tsnr} while both scenarios were initialized with the same input SNR and defined at the same spatial resolution. Furthermore, one can observe that the static refined strategy maintains a better tSNR than the dynamic refined one, but is associated instead with a more degraded image quality~(significant blurring). \Cref{fig:scenario3-tsnr} depicts the tSNR map for Scenario~3 under the \Ts{} model, which lies in the same range as for S2 while being smoother spatially. 

\newpage
\printbibliography{}
\end{document}